\documentclass[sn-aps,iicol]{sn-jnl} % Double column.

\graphicspath{{../fig/}{fig/}}
\newcommand{\key}[1]{\protect\includegraphics{key/#1}\,}
\usepackage{cleveref}
\crefname{figure}{Fig.}{Figs.}
\crefname{equation}{}{}
\crefname{section}{Section}{Sections}
\newcommand{\figbf}[1]{\textbf{#1}}
\newcommand{\figbfpic}[1]{\large\textbf{#1}}
\usepackage{overpic}
\jyear{2023}%

\begin{document}

\author[1]{\fnm{Petr} \sur{Karnakov}}\email{pkarnakov@seas.harvard.edu}

\author[2,1]{\fnm{Sergey} \sur{Litvinov}}\email{lisergey@ethz.ch}

\author*[1]{\fnm{Petros} \sur{Koumoutsakos}}\email{petros@seas.harvard.edu}

\affil[1]{\orgdiv{Computational Science and Engineering Laboratory},
\orgname{Harvard John A. Paulson School of Engineering and Applied Sciences}, \orgaddress{\street{29 Oxford St},
\city{Cambridge}, \postcode{02138}, \state{MA}, \country{United States}}}

\affil[2]{\orgdiv{Computational Science and Engineering Laboratory},
\orgname{ETH Zurich}, \orgaddress{\street{Clausiusstrasse 33}, \city{Zurich},
\postcode{8092}, \country{Switzerland}}}

\title[Flow reconstruction by multiresolution optimization]{%
  Flow reconstruction by multiresolution optimization of a discrete loss with automatic differentiation
}

\abstract{
We present a potent computational method for the solution of inverse problems
in fluid mechanics. We consider inverse problems formulated in terms of a
deterministic loss function that can accommodate data and regularization
terms. We introduce a multigrid decomposition technique that accelerates the
convergence of gradient-based methods for optimization problems with
parameters on a grid. We incorporate this multigrid technique to the ODIL
(Optimizing a DIscrete Loss) framework. The multiresolution ODIL (mODIL)
accelerates by an order of magnitude the original formalism and improves the
avoidance of local minima. Moreover, mODIL accommodates the use of automatic
differentiation for calculating the gradients of the loss function, thus
facilitating the implementation of the framework. We demonstrate the
capabilities of mODIL on a variety of inverse and flow reconstruction
problems: solution reconstruction for the Burgers equation, inferring
conductivity from temperature measurements, and inferring the body shape from
wake velocity measurements in three dimensions. We also provide a comparative
study with the related, popular Physics-Informed Neural Networks (PINNs)
method. We demonstrate that mODIL has three to five orders of magnitude
lower computational cost than PINNs in benchmark problems including simple PDEs and lid-driven cavity problems.
Our results suggest that mODIL is a very potent, fast and consistent method for solving inverse problems in fluid
mechanics.
}


\maketitle

\section{Introduction}

The domain of applications of  inverse problems spans 
many areas of science and engineering, including medical imaging,
geophysics, astronomy, materials science, and many others. We refer
the reader to the
books~\cite{kirsch2011introduction,aster2018parameter,banks2012estimation,engl1996regularization}.
Inverse problems play a crucial role in numerically solving partial differential equations, 
particularly in the field of fluid mechanics. 
A specific type of inverse problem in fluid mechanics is flow reconstruction, 
which involves estimating a flow field using limited measurements such as pressure or velocity. 
Flow reconstruction is an example of data assimilation, which involves combining mathematical 
models and observations to estimate the state of a system. 
The main challenge in both flow reconstruction and data assimilation is accurately 
estimating unobserved variables while accounting for uncertainties in the measurements and the model. 
Machine learning tools such as neural
networks have recently been used to solve these problems by
incorporating noisy data, solving differential equations, and
inferring unknown parameters and constitutive laws
\cite{Karniadakis21,he2020physics,wang2020towards,wang2022self}.
We argue that combining
traditional numerical methods with automatic differentiation and other machine learning tools  can solve these problems much faster. Data assimilation is often an ill-posed inverse problem because
the measurements obtained may only provide a limited and noisy
representation of the actual field due to factors such as
under-resolution. Weather forecasting is a common application of data
assimilation, where predictions are made based on unevenly distributed
data from weather stations worldwide~\cite{kalnay2003atmospheric}.

Solving inverse problems is challenging.
Inverse problems are often nonlinear, even when the
direct problem is linear. Moreover, inverse problems may not have a
unique solution so that small errors in measurements can cause
significant variations in the determined model. Finally, solving
inverse problems requires iterative techniques that involve solving
the direct problem multiple times, which can be computationally
expensive~\cite{kirsch2011introduction}.

We have proposed the  ODIL (Optimizing a Discrete Loss) framework~\cite{karnakov2022optimizing},
to address these challenges. First, ODIL is based on the 
discretization of the  forward problem, using modern machine learning
tools such as automatic differentiation to hide complexity and
non-linearity while maintaining its sparse structure. Additionally, if
the forward problem is linear and the resulting optimization problem
is quadratic, ODIL exploits this structure and can converge to an
exact solution in a single iteration.
Second, ODIL introduces a regularization term that encourages
smoothness, uniqueness, or stability of the solution, with few
limitations imposed on the term. This feature allows ODIL to apply to
situations where the forward problem may be ill-posed.

In this paper, in order to address the high computational cost of inverse problems,
we introduce multiresolution methods that exploit the multi-scale nature
of the forward problem by decomposing it into different scale bands,
each with different levels of detail. This decomposition allows the
optimization algorithm to focus on coarse-scale features of the
problem first and then refine the solution by adding finer-scale
details as needed. Starting with a coarse-scale approximation of the
solution and gradually refining, mODIL explores the
parameter space more effectively and avoids getting stuck in a local
minimum, resulting in faster convergence.
Finally, the corresponding optimization problem in ODIL has a standard
form and allows the use of popular optimization methods in machine
learning, from stochastic gradient descent (SGD) to more advanced
methods such as L-BFGS-B~\cite{zhu1997algorithm}.
As a result, ODIL can benefit from fast implementation on GPUs and
familiar programming tools such as TensorFlow and PyTorch.

In the following we  contrast ODIL with methods which are
implemented as end-to-end software solutions. A popular approach is
adjoint method, implemented in
\texttt{dolfin-adjoint}~\cite{mitusch2019dolfin}. It works by automatically
constructing and solving the adjoint equations associated with the
forward problem. The adjoint equations provide a way to efficiently
compute the derivatives of a quantity of interest with respect to the
input parameters, without requiring the user to manually derive and
implement the adjoint equations. However,  this approach is
limited to the situation where the forward and adjoint problem
are well-posed, while ODIL does not have this limitation. In addition,
\texttt{dolfin-adjoint} is a sophisticated tool that combines an open-source
platform for solving partial differential equations with a custom
automatic differentiation library. It requires a substantial amount of
expertise in mathematics and programming to use it effectively.
The nudging technique~\cite{di2020synchronization} adds a forcing term
to incorporate known velocity measurements
and also relies on a well-posed forward problem.
In the context of optimal control, the flow can be sampled
from the forward problem to obtain a policy via reinforcement learning~\cite{biferale2019zermelo}.

Another approach that has become popular recently is Physics-Informed
Neural Networks (PINNs) that use neural networks to represent unknown
fields and include a loss function that penalizes the mismatch between
predicted and observed data, as well as the right-hand side of the
differential equations, boundary, and initial
conditions~\cite{lagaris1998artificial, raissi2019physics,
Karniadakis21}. A key advantage of PINNs is the simplicity
of their implementation. However, PINNs have certain limitations
when it comes to addressing inverse problems: neural networks are
highly nonlinear functions and their approximation of the solution does not necessarily reflect the character of the PDEs (e.g. hyperbolic or parabolic cases. 
Furthermore PINNs do not fully exploit the
linearity inherent in the forward problem, leading to slow
convergence even for trivial cases.  Moreover, PINNs exhibit a spectral
bias, where solutions tend to be dominated by specific
modes~\cite{rahaman2019spectral,cao2019towards}. This bias arises
due to incomplete physical modeling, although efforts have been made
to design neural network architectures that capture the spectral
properties of the solution. PINNs are known for their slow convergence, which results
from the lack of sparse structure stemming from the locality of
physical laws. In contrast, ODIL reflects the character  of the underlying physical laws as it is based on consistent discretizations of the PDEs. Additionally, it can be difficult to harness the
multi-scale nature of the problem using PINNs. Finally, ODIL is
interpretable while PINNs are expressed by non-linear neural
networks. We remark that the grid-based discretization in ODIL may
suffer from the curse of dimensionality in high-dimensional
problems, while the error of PINNs typically scales as a square root
of the number of training points regardless of the space
dimensionality~\cite{mishra2022estimates}.

In this paper, we extend the ODIL
framework~\cite{karnakov2022optimizing} with a multigrid decomposition technique to accelerate the
convergence of gradient-based optimizers based on automatic differentiation.
We evaluate the technique on a series
of benchmark problems: Poisson's equation, solution reconstruction for the Burgers
equation, inferring a nonlinear conductivity from temperature measurements, and
inferring the body shape from velocity measurements. We demonstrate that the
mODIL technique reduces the number of iterations to achieve a certain error
by 10-100 times over ODIL. 
Furthermore, we show that it outperforms by orders of magnitude  PINNs.

\section{Methods}

\subsection{ODIL framework}

The (Optimizing a DIscrete Loss) ODIL framework formulates the problem as minimization of a loss function
that can include the residuals of the discretized governing equations,
terms to incorporate data, and regularization terms.

For example, consider a finite-difference discretization of 
the wave equation $u_{tt}=u_{xx}$ on a Cartesian grid in the space-time domain.
The following loss function
\begin{multline}
  \mathcal{L}(u)
  =
  \sum\limits_{(i,n)\in \Omega_1}
  \Big(
    \frac{u^{n+1}_i - 2u^n_i + u^{n-1}_i}{\Delta t^2} - \\
    \frac{u^n_{i+1} - 2u^n_i + u^n_{i-1}}{\Delta x^2}
  \Big)^2
  +
  \sum\limits_{(i,n)\in \Omega_2} \big( u^n_i - g^n_i \big)^2
\end{multline}
contains the residuals of the discretized equation in points $\Omega_1$
and terms to impose known values $g^n_i$ in points $\Omega_2$.
Here $u^n_i$ is a discrete field representing the solution.
This formulation covers all correct initial-value problems,
in which case $\Omega_1$ contains inner points,
$\Omega_2$ contains initial and boundary points,
the minimum is unique and the loss function evaluates to zero.
However, this formulation is more general
since points $\Omega_2$ can be placed anywhere in the space-time domain
and thus incorporate noisy or incomplete data.
The same idea extends to nonlinear equations.
To solve the minimization problem with a gradient-based method, such as
Adam~\cite{kingma2014adam} or L-BFGS-B~\cite{zhu1997algorithm},
we need the gradient of the loss, which computed using automatic differentiation
in TensorFlow~\cite{tensorflow2015whitepaper}.
To apply the Gauss-Newton method we linearize the discrete equations
to obtain a quadratic minimization problem and iteratively find the minimum
by solving a sparse linear system~\cite{karnakov2022optimizing}
with either a direct method~\cite{SciPyNMeth2020}
or an algebraic multigrid method~\cite{BeOlSc2022}.
For brevity, we refer to the above Gauss-Newton method as simply Newton's method
throughout the paper.
We use an implementation of L-BFGS-B from SciPy~\cite{SciPyNMeth2020}
and Adam from TensorFlow~\cite{tensorflow2015whitepaper}.

In our previous work~\cite{karnakov2022optimizing}, we compared ODIL
in terms of accuracy, convergence rate, and computational cost with
PINN~\cite{raissi2019physics} on a
set of forward and inverse problems for PDEs, showing that ODIL is
more computationally efficient than PINN by several orders of
magnitude.

\subsection{Multigrid decomposition}

Multigrid methods are generally accepted as
the fastest numerical methods for solving elliptic differential
equations~\cite{trottenberg2000multigrid}.
A standard multigrid method consists of the following parts:
a hierarchy of grids including the original grid and coarser grids,
discretizations of the problem on each grid,
interpolation operators to finer levels,
and restriction operators to coarser levels.
The method iteratively updates the solution on each level,
interpolates the update to finer levels,
and restricts the residuals to coarser levels.

As discussed in the previous section,
the optimization problem in ODIL can be solved with Newton's method
which involves a linear system at each step,
so the multigrid method can be applied directly to that linear system.
The conventional role of a multigrid method
is to act as a linear solver or a preconditioner for an iterative
method~\cite{siebenborn2017algorithmic,pinzon2022fluid,kothari2023multigrid,codd2018electrical,courty2006multilevel,akccelik2006parallel}.
Our previous results on ODIL~\cite{karnakov2022optimizing} has shown
that while Newton's method converges much faster than gradient-based methods,
it relies on sparsity of the linearized system and requires a linear solver
which determines the cost and may have limited efficiency especially on
GPUs~\cite{naumov2015amgx}.
Therefore, gradient-based optimizers
can become more efficient for larger problems or if the Jacobian matrix
with respect to certain parameters is dense, e.g. weights of a neural network.
However, all problems considered in~\cite{karnakov2022optimizing}
are solved faster by Newton's method than by gradient-based methods
due to their slow convergence.
The slow convergence is explained by the local nature of gradient-based methods.
An update in each grid point mostly depends on the gradient of the loss function
with respect to the value in that grid point, and possibly a limited history of
the gradients, e.g. momentum terms or an approximate inverse Hessian.

Here we propose a multigrid decomposition technique to accelerate the convergence of
optimization methods for problems that involve discrete fields on a grid.
Consider a uniform grid with $N_1=N$ points in each direction.
Introduce a hierarchy of successively coarser grids of size
$N_i = N / 2^{i-1}$ for $i=1,\dots,L$,
where $L$ is the total number of levels.
Define the multigrid decomposition operator as
\begin{multline}
M_L(u_1, \dots, u_L) = \\
  u_1 + T_1 u_2 + \dots +  T_1 T_2 \dots T_{L-1} u_{L},
\end{multline}
where each $u_i$ is a field on grid $N_i$,
and each $T_i$ is an interpolation operator from grid $N_{i+1}$ to
the finer grid~$N_i$.
The multigrid decomposition of a discrete field $u$ on a grid of size $N$ reads
\begin{equation}
u = M_L(u_1, \dots, u_L).
\end{equation}
Note that this representation is over-parameterized and therefore not unique.
The total number of scalar parameters increases from $N^d$ of the original field
$u$ to $N_1^d + \dots + N_L^d$ for the representation $u_1,\dots,u_L$.
The multigrid decomposition operator
can be implemented using Horner's scheme to reduce the number of
operator evaluations
\begin{multline}
  M_L(u_1, \dots, u_L) = \\
  u_1 + T_1 (u_2 + \dots  T_{L-2}(u_{L-1} + T_{L-1} u_{L})\dots).
\end{multline}

We define the interpolation operators~$T_i$ using linear
interpolation~\cite{trottenberg2000multigrid}.
We distinguish node-based and cell-based discretizations
on a uniform Cartesian grid consisting of $N^d$ cells in $d$ dimensions.
For a node-based discretization, the discrete field contains $N+1$ values in
each direction located in the grid nodes.
For a cell-based discretization, the discrete field contains $N$ values
in each direction located in the cell centers.
As an illustration, consider a hierarchy of grids in one dimension
with $N=N_1=8$, $N_2=4$, and $N_3=2$ cells.
The node-based interpolation matrices
$T_1\in\mathbb{R}^{9\times5}$ and $T_2\in\mathbb{R}^{5\times3}$ are
\begin{equation}
  T_1 =
  \frac{1}{2}
  \begin{bmatrix}
  2   & 0    &  0    &   0     & 0     \\
  1   & 1    &  0    &   0     & 0     \\
  0   & 2    &  0    &   0     & 0     \\
  0   & 1    &  1    &   0     & 0     \\
  0   & 0    &  2    &   0     & 0     \\
  0   & 0    &  1    &   1     & 0     \\
  0   & 0    &  0    &   2     & 0     \\
  0   & 0    &  0    &   1     & 1     \\
  0   & 0    &  0    &   0     & 2     \\
  \end{bmatrix}
  ,\quad
  T_2 =
  \frac{1}{2}
  \begin{bmatrix}
  2   & 0    &  0     \\
  1   & 1    &  0     \\
  0   & 2    &  0     \\
  0   & 1    &  1     \\
  0   & 0    &  2     \\
  \end{bmatrix}.
\end{equation}
The cell-based interpolation matrices
$T_1\in\mathbb{R}^{8\times4}$ and $T_2\in\mathbb{R}^{4\times2}$ are
\begin{equation}
  T_1 =
  \frac{1}{4}
  \begin{bmatrix}
  5 & -1 &  0    &   0    \\
  3 & 1  &  0    &   0    \\
  1 & 3  &  0    &   0    \\
  0    & 3  &  1 &   0    \\
  0    & 1  &  3 &   0    \\
  0    & 0     &  3 &   1 \\
  0    & 0     &  1 &   3 \\
  0    & 0     & -1 &   5 \\
  \end{bmatrix}
  ,\quad
  T_2 =
  \frac{1}{4}
  \begin{bmatrix}
  5  & -1 \\
  3  &  1 \\
  1 &  3 \\
 -1 &  5 \\
  \end{bmatrix}.
\end{equation}
Now we obtain mODIL by replacing a discrete field in ODIL with its multigrid decomposition.
The rest of the framework remains the same, including the discretized PDEs
and the optimization algorithm.
Gradients of the resulting loss function can be computed using automatic
differentiation.
This technique addresses the issue of locality of gradient-based optimizers
by extending the domain of dependence of each scalar parameter
so that information can propagate through the grid faster.

\section{Applications}
We demonstrate the acceleration offered by mODIL over ODIL and PINN in a number
of applications.
\subsection{Poisson equation}
\begin{figure*}
  \centering
  \begin{overpic}{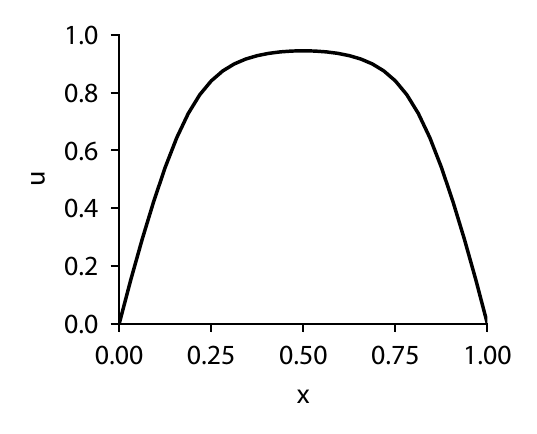}
    \put(5,75){\figbfpic{a}}
  \end{overpic}%
  \begin{overpic}{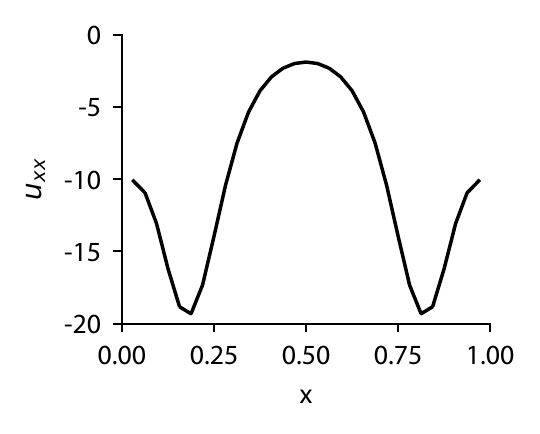}
    \put(5,75){\figbfpic{b}}
  \end{overpic}%
  \begin{overpic}{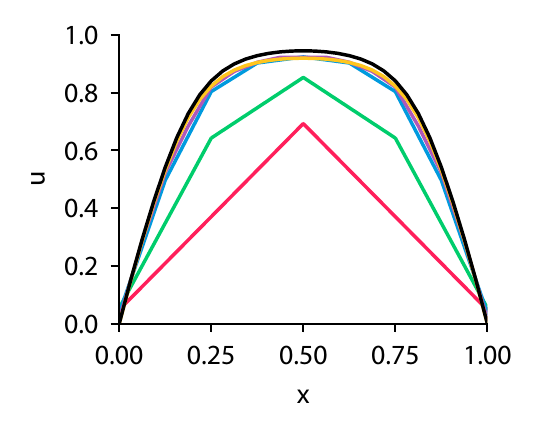}
    \put(5,75){\figbfpic{c}}
  \end{overpic}

  \begin{overpic}{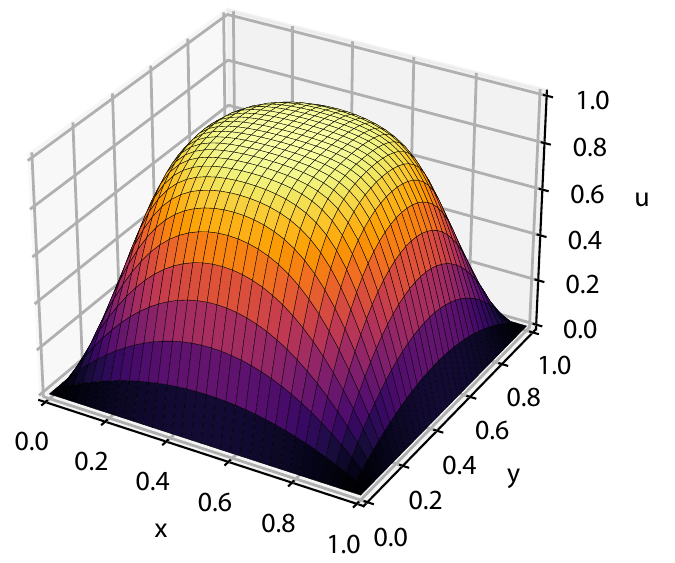}
    \put(8,75){\figbfpic{d}}
  \end{overpic}%
  \begin{overpic}{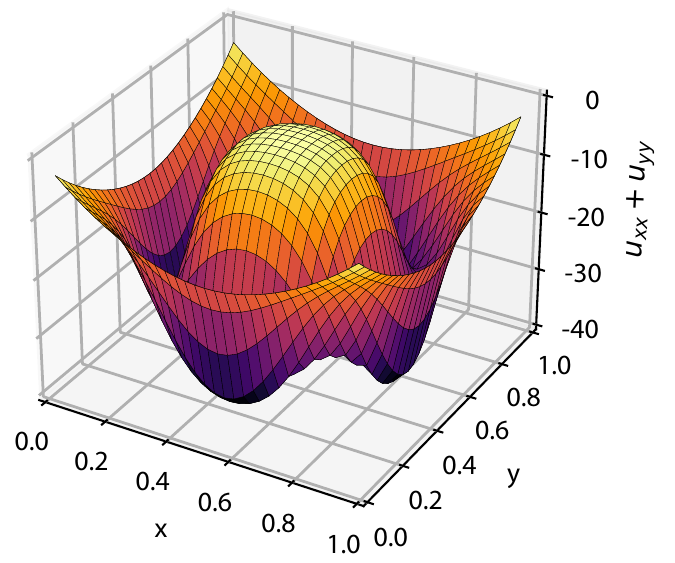}
    \put(8,75){\figbfpic{e}}
  \end{overpic}
  \caption{%
    Reference solution and the corresponding right-hand side of the Poisson
    equation.
    (\figbf{a},\figbf{b})~One-dimensional case $d=1$.
    (\figbf{c}) Cumulative sum of the multigrid levels after 400 iterations
    of Adam with node-based discretization
    compared to the reference solution~\key{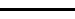}. The sum includes
    1~\key{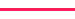}, 2~\key{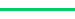}, 3~\key{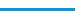}, 4~\key{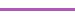}, and 5~\key{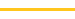}
    coarsest levels.
    (\figbf{d},\figbf{e})~Two-dimensional case $d=2$.
  }
  \label{f_pois_ref}
\end{figure*}

We solve the Poisson equation as a minimization problem to
study the effect of the multigrid decomposition
on the convergence rate of two optimization algorithms: Adam and L-BFGS-B.
As a benchmark problem, we choose a boundary value problem for the Poisson
equation with zero Dirichlet boundary conditions
\begin{equation}
  \begin{aligned}
    \nabla^2 u &= f, \quad \mathbf{x}\in\Omega, \\
    u &= 0, \quad\mathbf{x}\in\partial\Omega,
  \end{aligned}
\end{equation}
where $\Omega=[0,1]^d$ is a $d$-dimensional unit cube.
We consider two discretizations on a uniform Cartesian grid consisting of $N^d$
cells: a finite-volume discretization with cell-based values
and a finite-difference discretization with node-based values.
We define the reference solution as
\begin{equation}
  u_\mathrm{ref}(\mathbf{x}) = g\Big(\prod_{i=1}^d{5(1-x_i)x_i}\Big)
\end{equation}
where $g(v)=\big(v^5 / (1 + v^5)\big)^{1 / 5}$
and obtain the right-hand side $f(\mathbf{x})$
by evaluating the discretization of the Laplacian $\nabla^2$ on the reference
solution.
Therefore, the reference solution $u_\mathrm{ref}$ is the exact solution
of the discrete problem.
\Cref{f_pois_ref} shows the reference solution $u_\mathrm{ref}$ and the
corresponding right-hand side for 
the one- and two-dimensional ($d=1,2$) cases
with node-based values on a grid of size $N=32$ cells.
In the cell-based discretization, the discrete field
contains values in the cell centers and consists of $N$ values in each
direction and the Dirichlet boundary conditions are imposed using quadratic
extrapolation from the cell centers.
In the node-based discretization, the discrete field
contains values in the grid nodes and consists of $N+1$ values in each
direction and the Dirichlet boundary conditions are imposed directly
on the nodes.
Following the mODIL framework, we reformulate the problem as minimization
of a loss function
\begin{equation}
  \mathcal{L}(u)\approx \int_\Omega (\nabla^2 u - f)^2 {\rm d}V +
  \int_{\partial\Omega} u^2{\rm d}S
\end{equation}
and apply the multigrid decomposition to the unknown field.
The initial guess is zero.

\begin{figure*}
  \centering
  \begin{overpic}{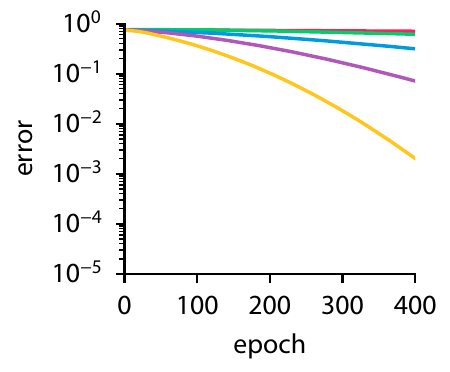}
    \put(5,80){\figbfpic{a}}
    \put(40,80){$d=1$, node}
  \end{overpic}%
  \begin{overpic}{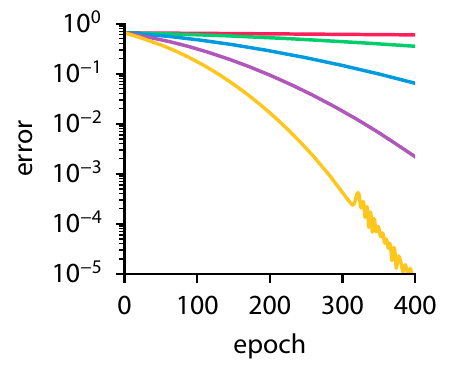}
    \put(5,80){\figbfpic{b}}
    \put(40,80){$d=2$, node}
  \end{overpic}%
  \begin{overpic}{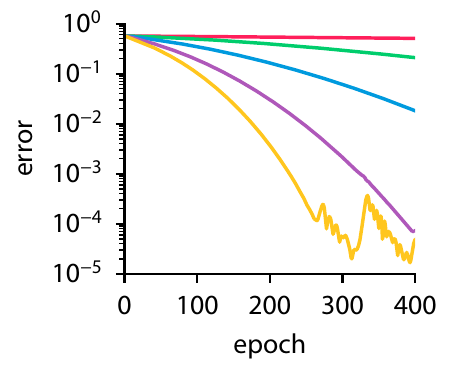}
    \put(5,80){\figbfpic{c}}
    \put(40,80){$d=3$, node}
  \end{overpic}
  \bigskip

  \begin{overpic}{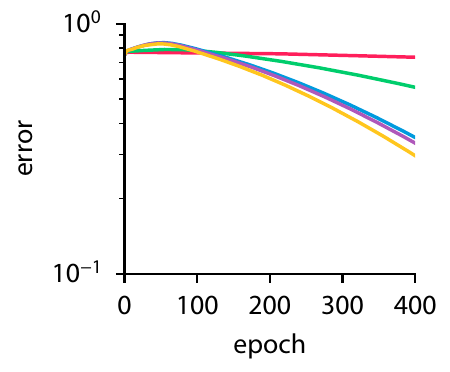}
    \put(5,80){\figbfpic{d}}
    \put(40,80){$d=1$, cell}
  \end{overpic}%
  \begin{overpic}{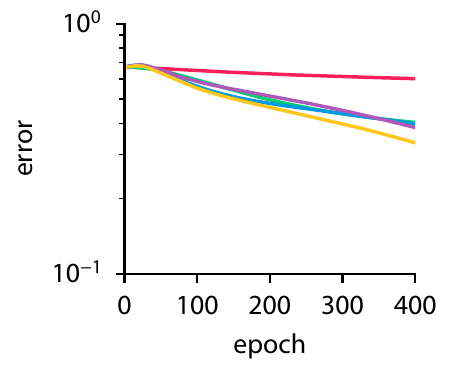}
    \put(5,80){\figbfpic{e}}
    \put(40,80){$d=2$, cell}
  \end{overpic}%
  \begin{overpic}{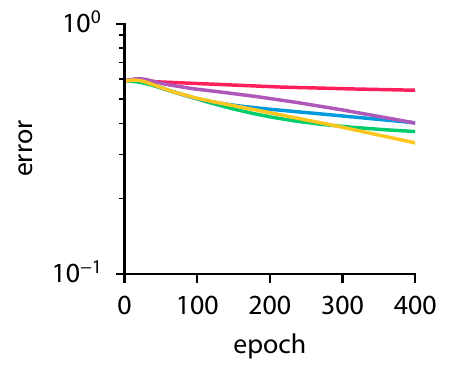}
    \put(5,80){\figbfpic{f}}
    \put(40,80){$d=3$, cell}
  \end{overpic}%
  \caption{%
    Convergence history of mODIL with Adam for the $d$-dimensional Poisson equation
    using $L=1$~\key{110}, 2~\key{120}, 3~\key{130}, 4~\key{140}, and 5~\key{150} levels.
    (\figbf{a},\figbf{b},\figbf{c})~Node-based discretization.
    (\figbf{d},\figbf{e},\figbf{f})~Cell-based discretization.
  }
  \label{f_pois_adam}
\end{figure*}

\Cref{f_pois_adam} shows the convergence history of mODIL with Adam
run for 400 iterations with the learning rate set to $0.005$.
The error is the root-mean-square (RMS) error relative to the exact solution.
The results include one-, two-, and three-dimensional cases ($d=1,2,3$)
both for the node-based and cell-based discretizations
using the multigrid decomposition with $L$ levels for $L=1$ to $5$.
The case of $L=1$ is equivalent to the original formulation on a single grid
of $32$ cells.
Cases with more levels $L>1$ include the original grid plus coarser levels
with 16, 8, 4, and 2 cells.
Overall, increasing the number of levels accelerates the convergence.
Also, the convergence is faster for the node-based discretization.
However, the convergence of Adam remains rather slow.
For the node-based discretization,
the error remains above $10^{-5}$ after 400 iterations.
For the cell-based discretization, the situation is even worse
with the error remaining above $0.2$.

\begin{figure*}
  \centering
  \begin{overpic}{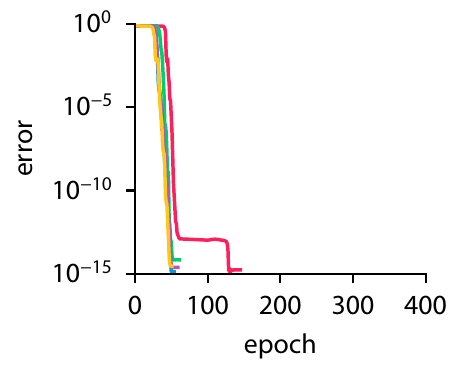}
    \put(5,80){\figbfpic{a}}
    \put(40,80){$d=1$, node}
  \end{overpic}%
  \begin{overpic}{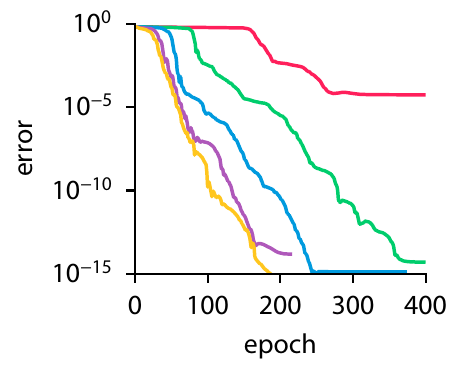}
    \put(5,80){\figbfpic{b}}
    \put(40,80){$d=2$, node}
  \end{overpic}%
  \begin{overpic}{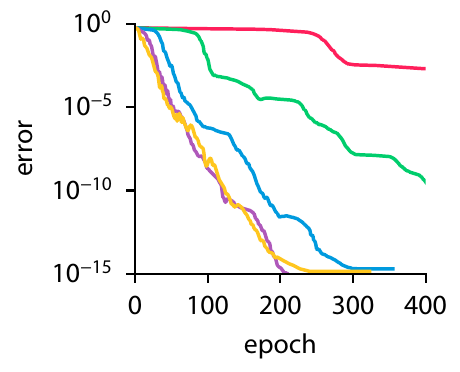}
    \put(5,80){\figbfpic{c}}
    \put(40,80){$d=3$, node}
  \end{overpic}
  \bigskip

  \begin{overpic}{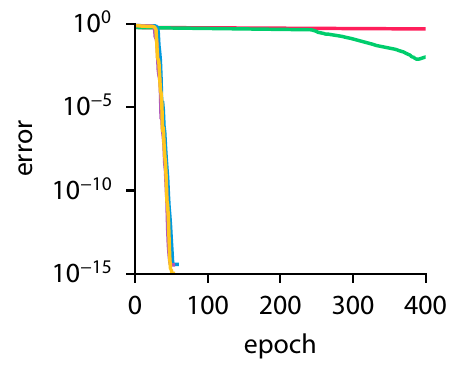}
    \put(5,80){\figbfpic{d}}
    \put(40,80){$d=1$, cell}
  \end{overpic}%
  \begin{overpic}{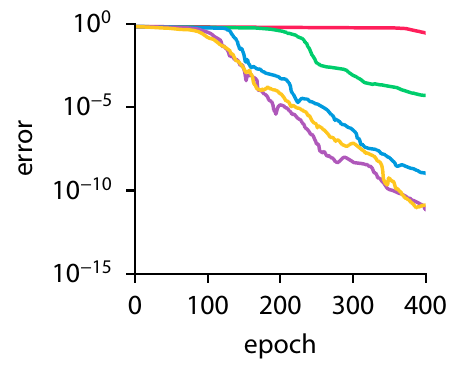}
    \put(5,80){\figbfpic{e}}
    \put(40,80){$d=2$, cell}
  \end{overpic}%
  \begin{overpic}{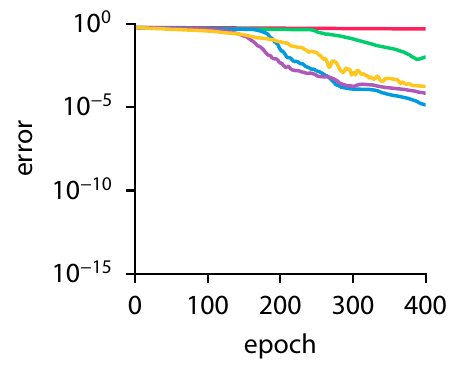}
    \put(5,80){\figbfpic{f}}
    \put(40,80){$d=3$, cell}
  \end{overpic}%
  \caption{%
    Convergence history of mODIL with L-BFGS-B for the $d$-dimensional Poisson equation
    using $L=1$~\key{110}, 2~\key{120}, 3~\key{130}, 4~\key{140}, and 5~\key{150} levels.
    (\figbf{a},\figbf{b},\figbf{c})~Node-based discretization.
    (\figbf{d},\figbf{e},\figbf{f})~Cell-based discretization.
  }
  \label{f_pois_bfgs}
\end{figure*}

\Cref{f_pois_bfgs} shows the convergence history of mODIL with L-BFGS-B
run for 400 iterations with the limited history of~50 vectors.
L-BFGS-B converges much faster than Adam.
In the one-dimensional case, the error achieves the machine precision
after about 50 iterations with three or more levels.
In the two- and three-dimensional cases, the error achieves the machine
precision after 200 iterations for the node-based discretization
using five levels.
Again, increasing the number of levels accelerates the convergence.

\subsection{Burgers equation: reconstruction}

Here we consider an ill-posed problem of reconstructing the solution
of the Burgers equation from sparse measurements.
We solve the problem in a unit domain $(x,t)\in[0,1]^2$.
The problem is to find a solution $u(x,t)$ to the Burgers equation
$u_t + uu_x=0$ that takes known values $u(x_i,t_i)=u_i$
in a finite set of measurement points $(x_i,t_i)$ for $i=1,\dots,N$.
In this example, we imposed the reference solution in 84 points
placed on the edges of a rectangle.
We use a finite volume discretization on a uniform grid
using a first-order upwind scheme for the flux.
The loss function is a discretization of the squared residual
and terms to impose the known values
\begin{equation}
  \mathcal{L}(u)\approx \int (u_t + uu_x)^2 {\rm d}V
  + \frac{1}{N}\sum_{i=1}^N(u(x_i,t_i)-u_i)^2
\end{equation}
The grid consist of $64\times64$ cells.
To generate the reference solution, we solve the discrete problem
with zero Dirichlet boundary conditions
and the initial condition $u=(1 - \cos{6\pi x}) / 2$.
In the case of ODIL optimized with Newton's method,
we amend the loss function with regularization terms and obtain
\begin{equation}
\begin{aligned}
  \mathcal{L}(u)&\approx \int (u_t + uu_x)^2 {\rm d}V
  + \frac{1}{N}\sum_{i=1}^N(u(x_i,t_i)-u_i)^2 \\
  &+ k_\mathrm{xreg} \int u_x^2{\rm dV} + k_\mathrm{treg} \int u_t^2{\rm dV}
\end{aligned}
\end{equation}
The regularization coefficients decay with iterations and take values
$k_\mathrm{xreg} = k_\mathrm{treg} = 0.01 \cdot 2^{-n/3}$,
where $n$ is the iteration number.
This regularization enables convergence of Newton's method.

\begin{figure*}
  \centering
  \begin{minipage}[c]{7.7cm}
    \hspace{-12mm}
    \begin{overpic}{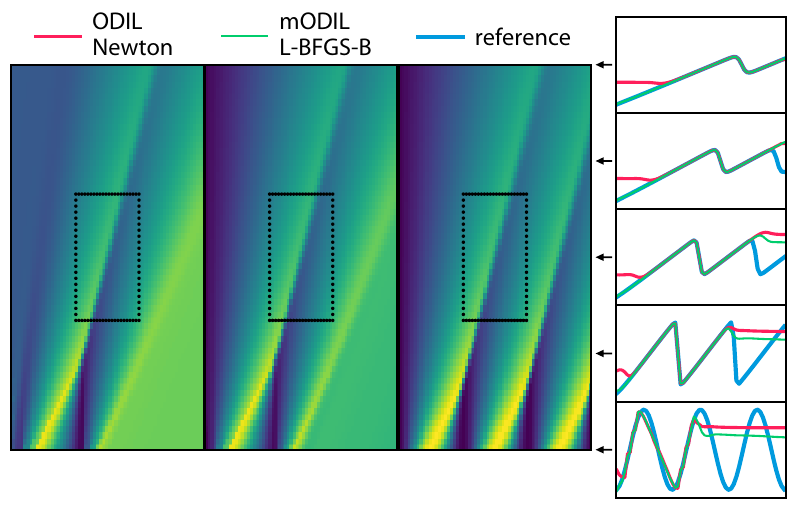}
      \put(-3,58){\figbfpic{a}}
      \put(-3,20){\rotatebox{90}{time \raisebox{0.5pt}{\protect\includegraphics[width=12mm]{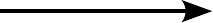}}}}
    \end{overpic}
  \end{minipage}
  \begin{minipage}[c]{2.7cm}
    \vspace{-5.3mm}
    \begin{overpic}{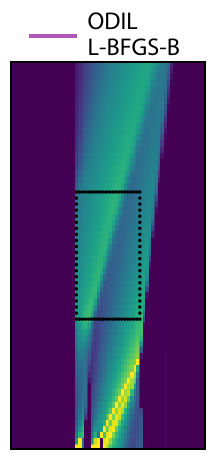}
      \put(-6,88.5){\figbfpic{b}}
    \end{overpic}
  \end{minipage}
  \begin{minipage}[c]{3cm}
    \begin{overpic}{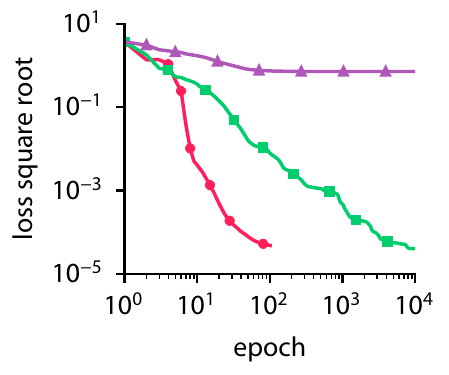}
      \put(0,78){\figbfpic{c}}
    \end{overpic}

    \begin{overpic}{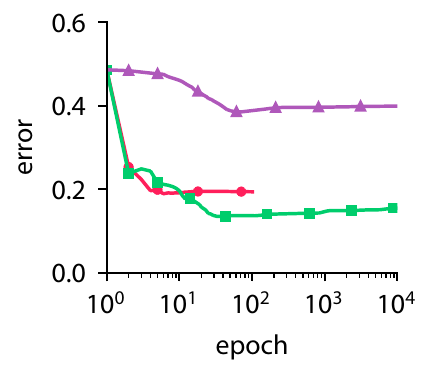}
      \put(0,78){\figbfpic{d}}
    \end{overpic}
  \end{minipage}
  \caption{%
    Reconstructing the solution of the Burgers equation from sparse measurements.
    (\figbf{a})~Solution found by ODIL with Newton and mODIL with L-BFGS-B
    compared to the reference solution.
    The measurement points are on the perimeter of a rectangle (black dots).
    (\figbf{b})~Solution found by ODIL with L-BFGS-B.
    (\figbf{c},\figbf{d})~Convergence history of 
    ODIL with Newton~\key{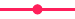}, mODIL with L-BFGS-B~\key{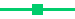},
    and ODIL with L-BFGS-B~\key{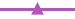}
    showing the square root of the loss function
    and the RMS error relative to the reference solution.
  }
  \label{f_burg}
\end{figure*}

\Cref{f_burg} shows the obtained solutions in the space-time domain
and the convergence history of various optimization methods:
ODIL with Newton, mODIL with L-BFGS-B using six levels (64, 32, 16, 8, 4, and 2 cells),
and ODIL with L-BFGS-B.
The error is the RMS error relative to the reference solution.
We note that the error does not converge to zero since
imposing the reference solution on the edges of a rectangle
is not sufficient to extend the solution throughout the domain.
However, all optimization methods recover the solution inside the rectangle.
ODIL with Newton demonstrates the fastest convergence, reaching an error
of 0.2 after about 10 iterations.
mODIL with L-BFGS-B reaches an error of 0.15 after about 40 iterations.
Both methods produce a solution that is consistent with the imposed
solution in the areas spanned by the characteristics extending from the
rectangle. The loss function achieves values below $10^{-4}$.
Conversely, ODIL with L-BFGS-B stops at a larger error of 0.4
even after 10\,000 iterations and produces a qualitatively different solution
that appears to be a local minimum.

\begin{figure*}
  \centering
  \begin{minipage}[c]{7.7cm}
    \hspace{-12mm}
    \begin{overpic}{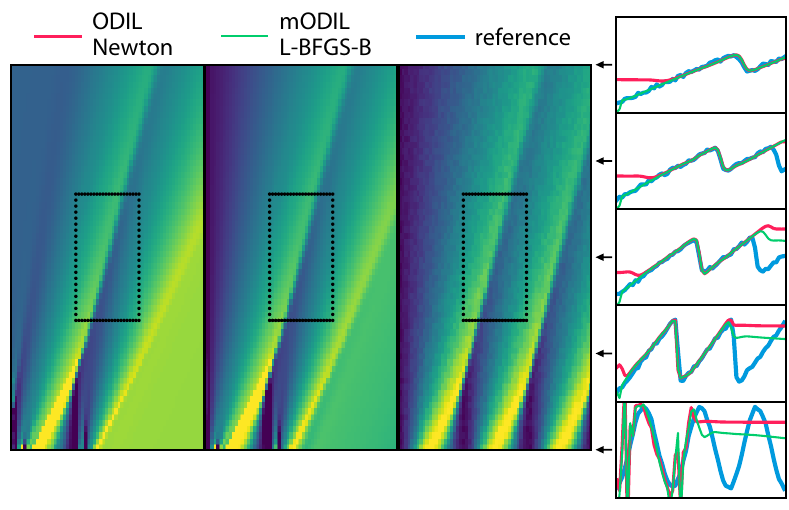}
      \put(-3,58){\figbfpic{a}}
      \put(-3,20){\rotatebox{90}{time \raisebox{0.5pt}{\protect\includegraphics[width=12mm]{arrow.pdf}}}}
    \end{overpic}
  \end{minipage}
  \begin{minipage}[c]{2.7cm}
    \vspace{-5.3mm}
    \begin{overpic}{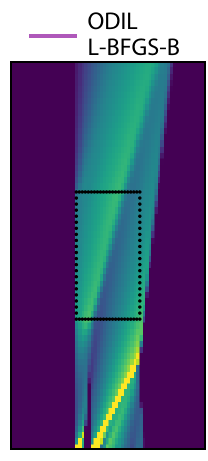}
      \put(-6,88.5){\figbfpic{b}}
    \end{overpic}
  \end{minipage}
  \begin{minipage}[c]{3cm}
    \begin{overpic}{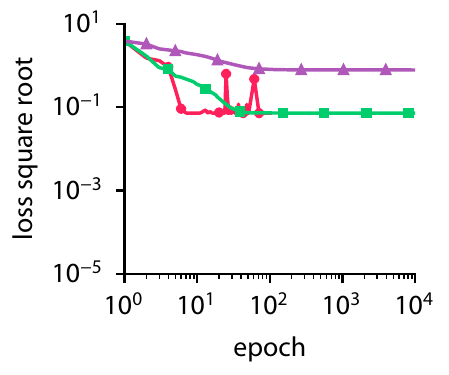}
      \put(0,78){\figbfpic{c}}
    \end{overpic}

    \begin{overpic}{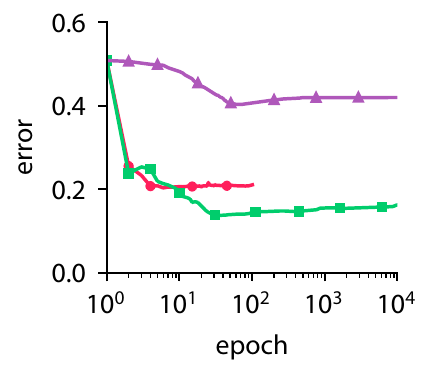}
      \put(0,78){\figbfpic{d}}
    \end{overpic}
  \end{minipage}
  \caption{%
    Reconstructing the solution of the Burgers equation from sparse measurements.
    The reference solution is perturbed by uniform noise.
    (\figbf{a})~Solution found by ODIL with Newton and mODIL with L-BFGS-B
    compared to the reference solution.
    The measurement points are on the perimeter of a rectangle (black dots).
    (\figbf{b})~Solution found by ODIL with L-BFGS-B.
    (\figbf{c},\figbf{d})~Convergence history of 
    ODIL with Newton~\key{111}, mODIL with L-BFGS-B~\key{122},
    and ODIL with L-BFGS-B~\key{144}
    showing the square root of the loss function
    and the RMS error relative to the reference solution.
  }
  \label{f_burg_noise}
\end{figure*}

To evaluate the method on noisy data, we
add uniform noise $U[0,0.05]$ to the reference solution and repeat the analysis.
\Cref{f_burg_noise} shows the results.
The error behaves similar to the case without the noise.
The loss function now only reaches values below $0.1$,
since the imposed data points are no longer consistent with a solution
of the Burgers equation.
Also, the inferred solution has rapid changes near the initial time $t=0$.
Again, ODIL with L-BFGS-B produces a solution that is qualitatively different and has a larger error.

\subsection{Lid-driven cavity}
\label{s_lidforw}

The lid-driven cavity problem is a standard test case~\cite{ghia1982high} for
numerical methods for the steady-state Navier-Stokes equations in two
dimensions
\begin{equation}
  \begin{aligned}
    u_x + v_y &= 0, \\
    u u_x + v u_y &= -p_x  + 1/\mathrm{Re} (u_{xx} + u_{yy}), \\
    u v_x + v v_y &= -p_y  + 1/\mathrm{Re} (v_{xx} + v_{yy}),
  \end{aligned}
\end{equation}
where $u(x,y)$ and $v(x,y)$ are the two velocity components and $p(x,y)$ is the
pressure.
The problem is solved in a unit domain with no-slip boundary conditions.
The upper boundary is moving to the right at a unit velocity
while the other boundaries are stagnant.
We apply both mODIL and PINN to this problem.
To represent the solution in mODIL, we use a uniform grid of $65\times 65$ cells
with the multigrid decomposition.
We use a finite volume discretization on a uniform Cartesian grid
based on the SIMPLE method~\cite{patankar1983,ferziger2012}
with the Rhie-Chow interpolation~\cite{rhie1983numerical}
to prevent oscillations in the pressure field
and the deferred correction approach
that treats high-order discretization explicitly
and low-order discretization implicitly to obtain an operator with a compact
stencil.
To represent the solution $(u,v,p)$ in PINN, we use a fully-connected
neural network of size $2\times 32\times 32\times 32\times 3$.
The number of collocation points for PINN is 10\,000 points inside the domain
and 400 for the boundary conditions.
We use L-BFGS-B to solve the optimization problem for both methods.

\begin{figure*}
  \centering
  \begin{minipage}[c]{3cm}
    \vspace{-4mm}
    \begin{overpic}[height=30mm]{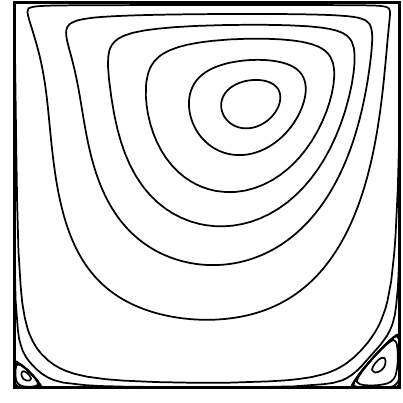}
      \put(5,105){\figbfpic{a}}
      \put(20,102){\raisebox{0.5pt}{\protect\includegraphics[width=18mm]{arrow.pdf}}}
      \put(30,-10){$\mathrm{Re}=100$}
    \end{overpic}
  \end{minipage}\hspace{1cm}
  \begin{minipage}[c]{5cm}
    \begin{overpic}[]{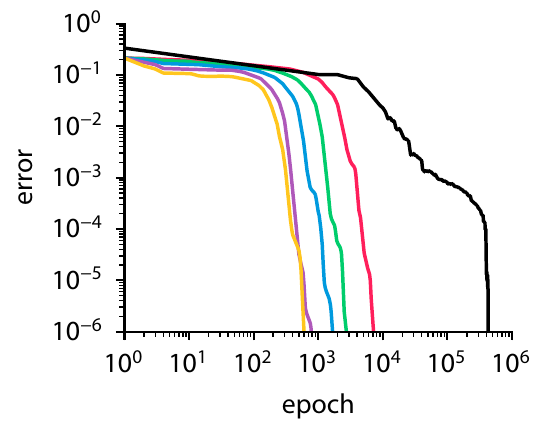}
      \put(1,74){\figbfpic{b}}
    \end{overpic}
  \end{minipage}

  \begin{minipage}[c]{\textwidth}
    \vspace{4mm}
    \centering
    \begin{overpic}[width=2.5cm]{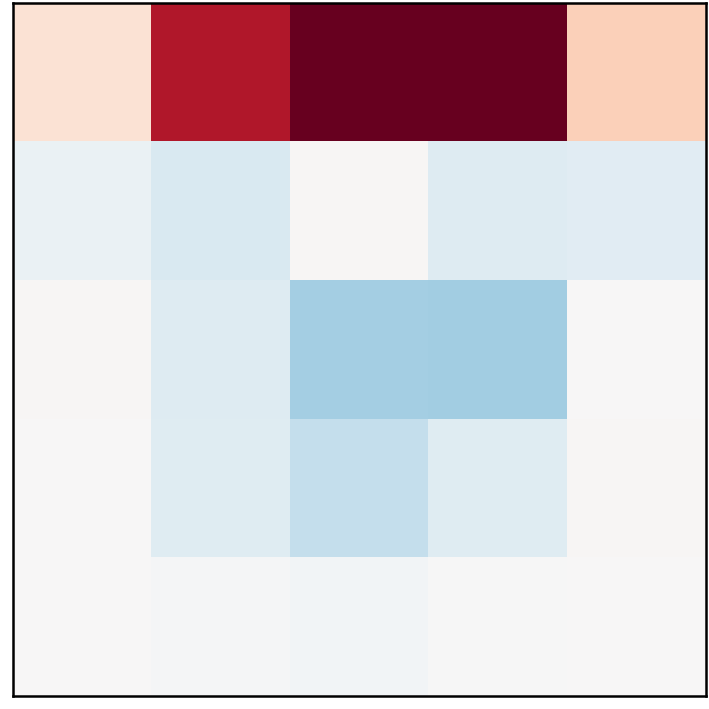}
      \put(1,105){\figbfpic{c}}
      \put(30,105){$5\times 5$}
    \end{overpic}
    \begin{overpic}[width=2.5cm]{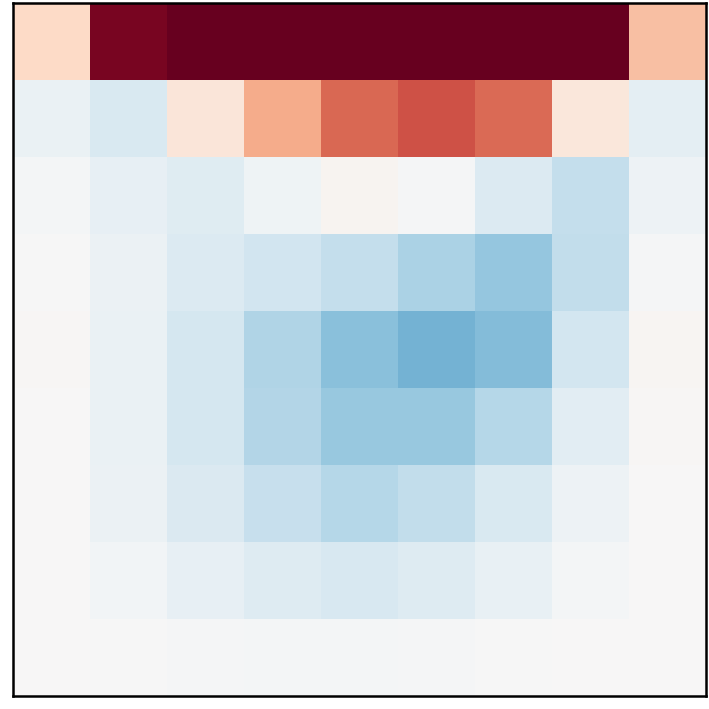}
      \put(1,105){\figbfpic{d}}
      \put(30,105){$9\times 9$}
    \end{overpic}
    \begin{overpic}[width=2.5cm]{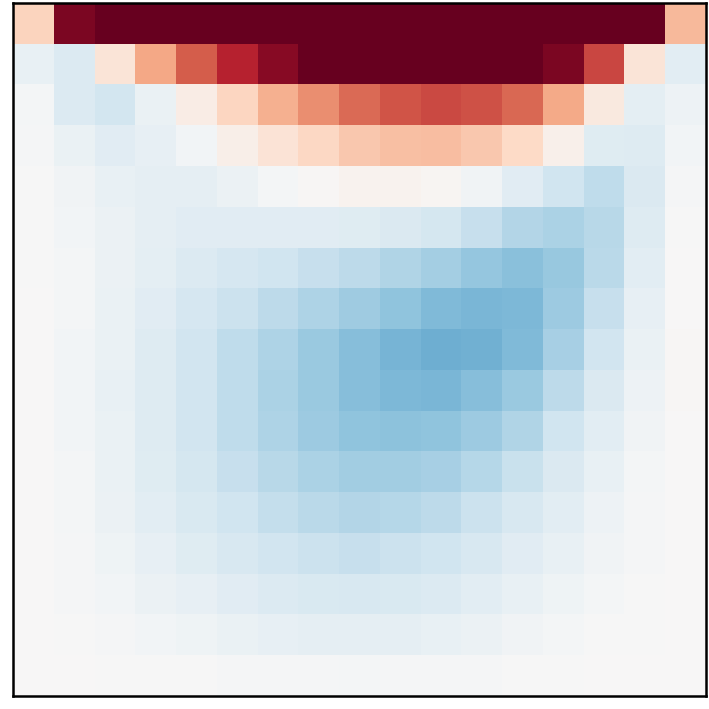}
      \put(1,105){\figbfpic{e}}
      \put(30,105){$17\times 17$}
    \end{overpic}
    \begin{overpic}[width=2.5cm]{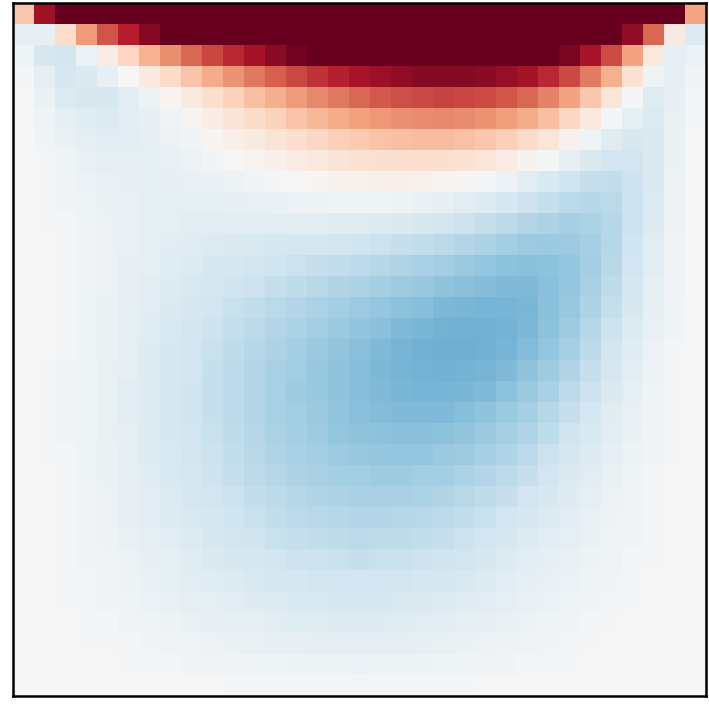}
      \put(1,105){\figbfpic{f}}
      \put(30,105){$33\times 33$}
    \end{overpic}
    \begin{overpic}[width=2.5cm]{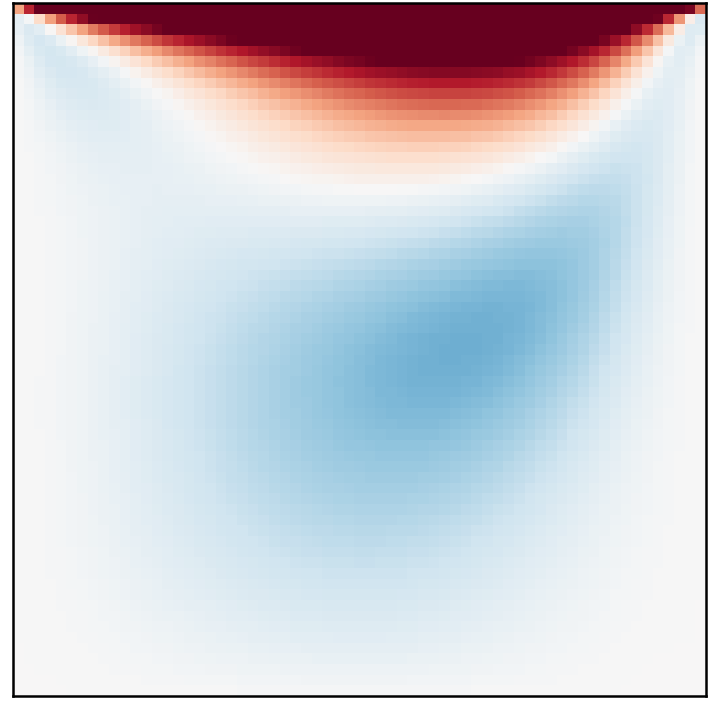}
      \put(1,105){\figbfpic{g}}
      \put(30,105){$65\times 65$}
    \end{overpic}
  \end{minipage}

  \caption{%
    Lid-driven cavity flow at $\mathrm{Re}=100$ solved using mODIL and PINN.
    (\figbf{a})~Streamlines from mODIL. The top wall moves along the arrow.
    (\figbf{b})~Convergence history of mODIL with L-BFGS-B
    using $L=1$~\key{110}, 2~\key{120}, 3~\key{130}, 4~\key{140}, and 5~\key{150}
    levels, and PINN~\key{100}
    showing the RMS error in velocity~$u$ relative to the final result of each method.
    (\figbf{c},\figbf{d},\figbf{e},\figbf{f},\figbf{g})~Multigrid levels of velocity $u$ obtained by mODIL:
    $u_5$ (\figbf{c}),
    $u_4+T_4 u_5$ (\figbf{d}),
    $u_3+T_3(u_4 + T_4 u_5)$ (\figbf{e}),
    $u_2 + T_2(u_3+T_3(u_4 + T_4 u_5))$ (\figbf{f}),
    and $u=u_1 +T_1(u_2 + T_2(u_3+T_3(u_4 + T_4 u_5)))$ (\figbf{g})
    with interpolation operators $T_1,T_2,T_3,T_4$.
  }
  \label{f_cavity_mg}
\end{figure*}

\Cref{f_cavity_mg} shows the streamlines at $\mathrm{Re}=100$ obtained using mODIL,
as well as a convergence history of L-BFGS-B depending on the number of multigrid levels.
The RMS error in velocity $u$ is computed relative to the solution
of the discrete problem in the case of mODIL
or the solution at iteration 420\,000 in the case of PINN.
mODIL with $L=5$ levels (65, 33, 17, 9, and 5 cells)
shows the fastest convergence,
taking 320 iterations to reach an error of $10^{-3}$.
ODIL (equivalent to mODIL with $L=1$) takes 3840 iterations.
PINN takes 70\,000 iterations to reach the same error,
which is 20x more than ODIL and 200x more than mODIL with $L=5$.

\subsection{Inferring conductivity from temperature}
\label{s_heat}

Here we consider an inverse problem of inferring a conductivity
function from temperature measurements.
We solve the problem in a unit domain $(x,t)\in[0,1]^2$.
The problem is to find a nonlinear conductivity function $k(u)$
and temperature field $u(x,t)$ that satisfies the heat equation
$u_t - (k(u)u_x)_x=0$ with zero Dirichlet boundary conditions $u(0,t)=u(1,t)=0$
and initial conditions $u(x,0)=g(x) - g(0)$,
where $g(x) = e^{-50(x - 0.5)^2}$.
In addition, the temperature field needs to take known values $u(x_i,t_i)=u_i$
in a finite set of measurement points $(x_i,t_i)$ for $i=1,\dots,500$.
We discretize the equation on a uniform grid with cell-based values as
\begin{multline}
\frac{u^{n+1}_i-u^n_i}{\Delta t} -\\
\frac{k(u_{i+1/2})(u^{n+1}_{i+1}-u^{n+1}_{i})
  - k(u_{i-1/2})(u^{n+1}_{i}-u^{n+1}_{i-1})}{\Delta h^2} \\
  = 0
\end{multline}
where $u_{i+1/2}=(u^{n+1}_{i+1} + u^{n+1}_{i})/2$.
The loss function for the inverse problem consists of the residuals of the
equation and the quadratic terms to impose the temperature values
as well as the initial and boundary conditions.

\begin{figure*}
  \centering
  \begin{overpic}[]{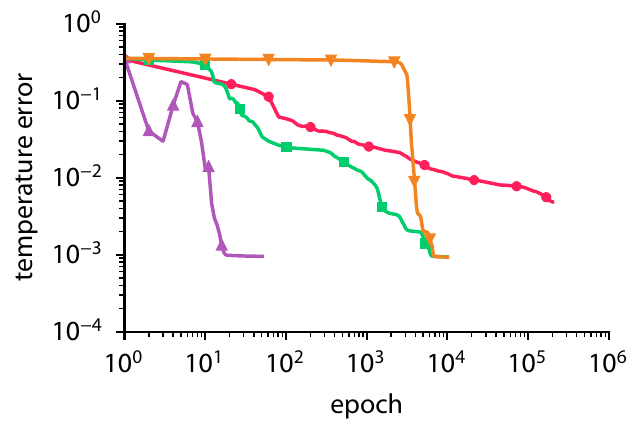}
    \put(4,60){\figbfpic{a}}
  \end{overpic}%
  \begin{overpic}[]{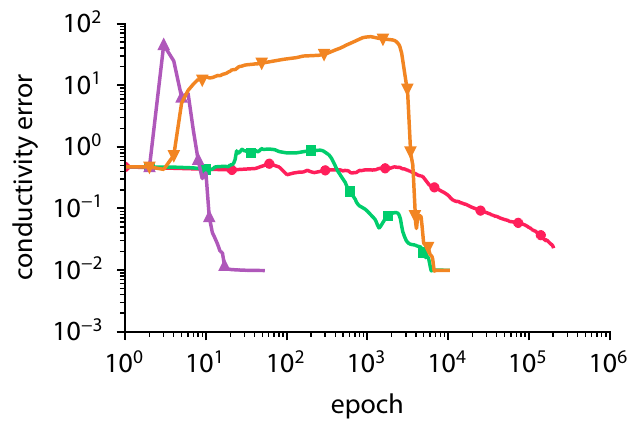}
    \put(3,60){\figbfpic{b}}
  \end{overpic}

  \centering
  \begin{overpic}[]{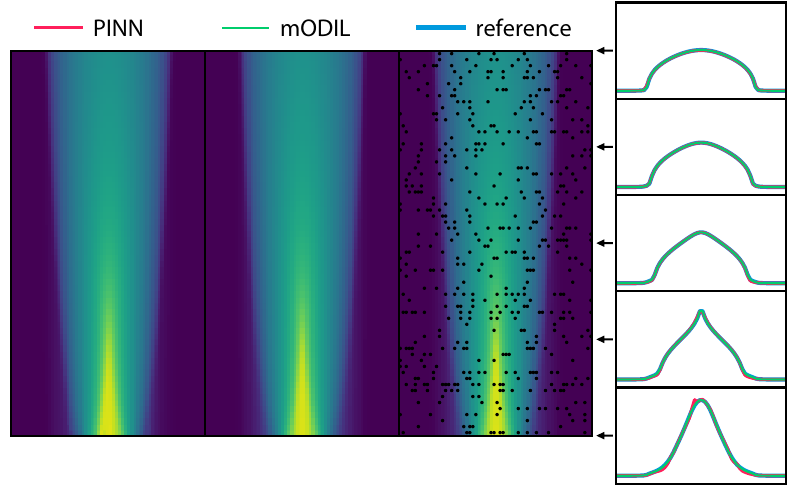}
    \put(-4,55){\figbfpic{c}}
    \put(-3,20){\rotatebox{90}{time \raisebox{0.5pt}{\protect\includegraphics[width=12mm]{arrow.pdf}}}}
  \end{overpic}\qquad
  \begin{overpic}[]{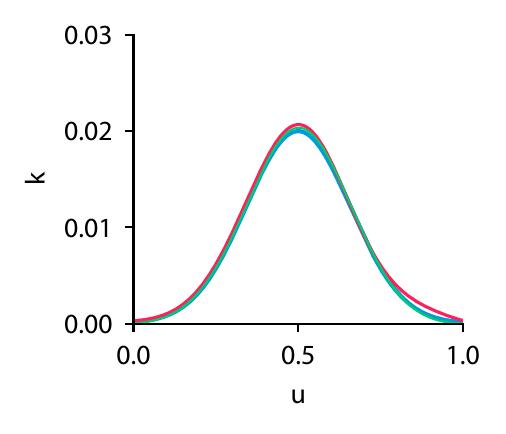}
    \put(5,80){\figbfpic{d}}
  \end{overpic}

  \caption{%
    Inferring conductivity from temperature measurements.
    (\figbf{a},\figbf{b}) Convergence history of
    PINN with L-BFGS-B~\key{111},
    mODIL with L-BFGS-B~\key{122},
    ODIL with L-BFGS-B~\key{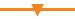},
    ODIL with Newton~\key{144},
    showing the RMS error in temperature
    and conductivity normalized by its maximum reference value.
    (\figbf{c},\figbf{d}) Inferred temperature field and conductivity function
    from PINN~\key{110} and ODIL~\key{120}
    compared to the reference solution~\key{130}.
    The measurement points are shown in the reference temperature field (black dots).
  }
  \label{f_heat}
\end{figure*}

To generate the temperature measurements and the reference solution,
we specify the conductivity function
as $k(u)=0.02\,e^{-20(u - 0.5)^2}$ and solve the forward problem on a grid of
$256\times 256$ cells.
Then we solve the inverse problem using PINN and ODIL and compare the results.
To represent the unknown conductivity function $k(u)$, we use a fully-connected
neural network of size $1\times 5\times 5\times 1$,
i.e. one input $u$, two hidden layers with five neurons in each layer, and one
output $k$, which is then squared to ensure that the conductivity is
non-negative.
To represent the temperature field $u(x,t)$, we use
a fully-connected neural network of size $2\times 32\times 32\times 32\times
32\times 1$ in PINN and a uniform grid of $64\times 64$ cells in ODIL.
The number of collocation points for PINN is 4096 points inside the domain
and 384 for the initial and boundary conditions.
\Cref{f_heat} shows the convergence history, the inferred temperature and conductivity.
In the case of ODIL optimized with Newton's method,
we add a regularization term $k_w^2\|w - w^*\|_2^2$ for the weights of the
neural network $k(u)$, where $k_w=0.8$ is a parameter,
$w$ is a vector of all weights, and
$w^*$ is a vector with the same weights but ``frozen'' so they are ignored
in the linearization of the problem.
This regularization introduces damping for the weights
but does not affect the solution if the method converges.
Both PINN and ODIL infer similar temperature fields and conductivity functions.
The convergence history includes the RMS error in the temperature
field relative to the reference solution
and the RMS error in the conductivity function
in the range $u\in(0,1)$ with both quantities normalized by their maximum values
in the reference solution.
ODIL with Newton demonstrates the fastest convergence, which
only takes 11 iterations to achieve an error of 5\% for the conductivity
function.
ODIL with L-BFGS-B takes 4000 iterations to achieve the same error.
Then, mODIL converges faster than ODIL, taking 1500 iterations, 
and in more regular way compared to ODIL,
e.g. the conductivity error monotonically decreases while ODIL passes through
solutions with a rapidly increasing error.
Therefore, the multigrid decomposition regularizes the solution.
In contrast, PINN converges significantly slower, achieving an error 5\% after
100\,000 iterations.
The corresponding execution time on one CPU core
amounts to about 10 hours for PINN,
125~s for ODIL with L-BFGS-B,
and 120~s for ODIL with Newton.
Consistent with our previous observations,
ODIL takes fewer iterations than PINN and each iteration is cheaper,
which results in two-three orders of magnitude lower computational cost overall.

\subsection{Inferring body shape from velocity}

Here we consider a three-dimensional inverse problem of inferring the shape of
a body from measurements of the flow velocity around the body.
The model consists of the steady-state Navier-Stokes equations
with penalization terms
to impose the no-slip conditions on the body~\cite{angot1999penalization}
\begin{equation}
  \begin{aligned}
    \nabla \cdot \mathbf{u} &= 0, \\
    (1-\chi) \big((\mathbf{u}\cdot\nabla) \mathbf{u} +\nabla p  -
    \tfrac{D}{\mathrm{Re}} \nabla^2\mathbf{u}\big)
    + \lambda \chi \mathbf{u} &=0, \\
  \end{aligned}
  \label{eq_nschi3}
\end{equation}
where $\lambda$ is a penalization parameter and $D$ is a characteristic length of the body.
The shape of the body is described by the body fraction~$\chi(\mathbf{x})$
which takes values $\chi=1$ inside the body and $\chi=0$ outside.
The problem is solved in the domain $[0,2]\times[0,1]\times[0,1]$ with the inlet condition
$\mathbf{u}=(1, 0, 0)$ at $x=0$, outlet condition $p=0$ at $x=2$, and free-slip
walls on the other boundaries. This setup describes the flow past a body in a
channel confined by free-slip walls.
The discretization follows~\cref{s_lidforw}.
The forward problem is to find the velocity $\mathbf{u}$ and pressure $p$
that satisfy equations~\cref{eq_nschi3} given a prescribed body fraction~$\chi$.
The inverse problem is to find the velocity $\mathbf{u}$, pressure $p$, and body
fraction~$\chi$ that satisfy equations~\cref{eq_nschi3}
such that the velocity field takes known values
$\mathbf{u}(\mathbf{x}_i)=\mathbf{u}_i$
in a finite set of $N$ measurement points $\mathbf{x}_i$ for $i=1,\dots,N$.

To solve the inverse problem using ODIL,
we formulate it as minimization of the loss function in terms of the unknown fields:
velocity $\mathbf{u}$, pressure $p$, and body fraction $\hat\chi$.
Here $\hat\chi$ is a transformed body fraction
defined as $\chi = 1 / (1 + e^{-(\hat\chi+5)})$,
so that during the optimization the body fraction~$\chi$ only takes values between~0 and~1.
The loss function is a sum of the residuals of equations~\cref{eq_nschi3}
and terms to impose the reference data.
The penalization parameter is set to $\lambda=1$.
The problem is solved on a $129\times65\times65$ grid and
the reference data is obtained from the forward problem.
The characteristic length of the body is taken to be $D=0.4$
and the Reynolds number is $\mathrm{Re}=60$.
To solve the optimization problem, we use L-BFGS~\cite{liu1989limited}
implemented in TensorFlow Probability~\cite{dillon2017tensorflow}.
To accelerate the convergence, we apply the multigrid decomposition with $L=6$ levels.
The initial guess is $\mathbf{u}=(1, 0, 0)$ for the velocity,
$p=0$ for the pressure, and $\hat\chi=0$ for the transformed body fraction.
According to the above transformation, the corresponding initial guess for
the body fraction is $\chi=1 / (1 + e^{5})$.
We terminate the algorithm after 10\,000 epochs for the forward problem
and 20\,000 epochs for the inverse problem.

\begin{figure*}
  \centering
  \begin{minipage}[t]{0.4\textwidth}
    \begin{overpic}[width=7cm]{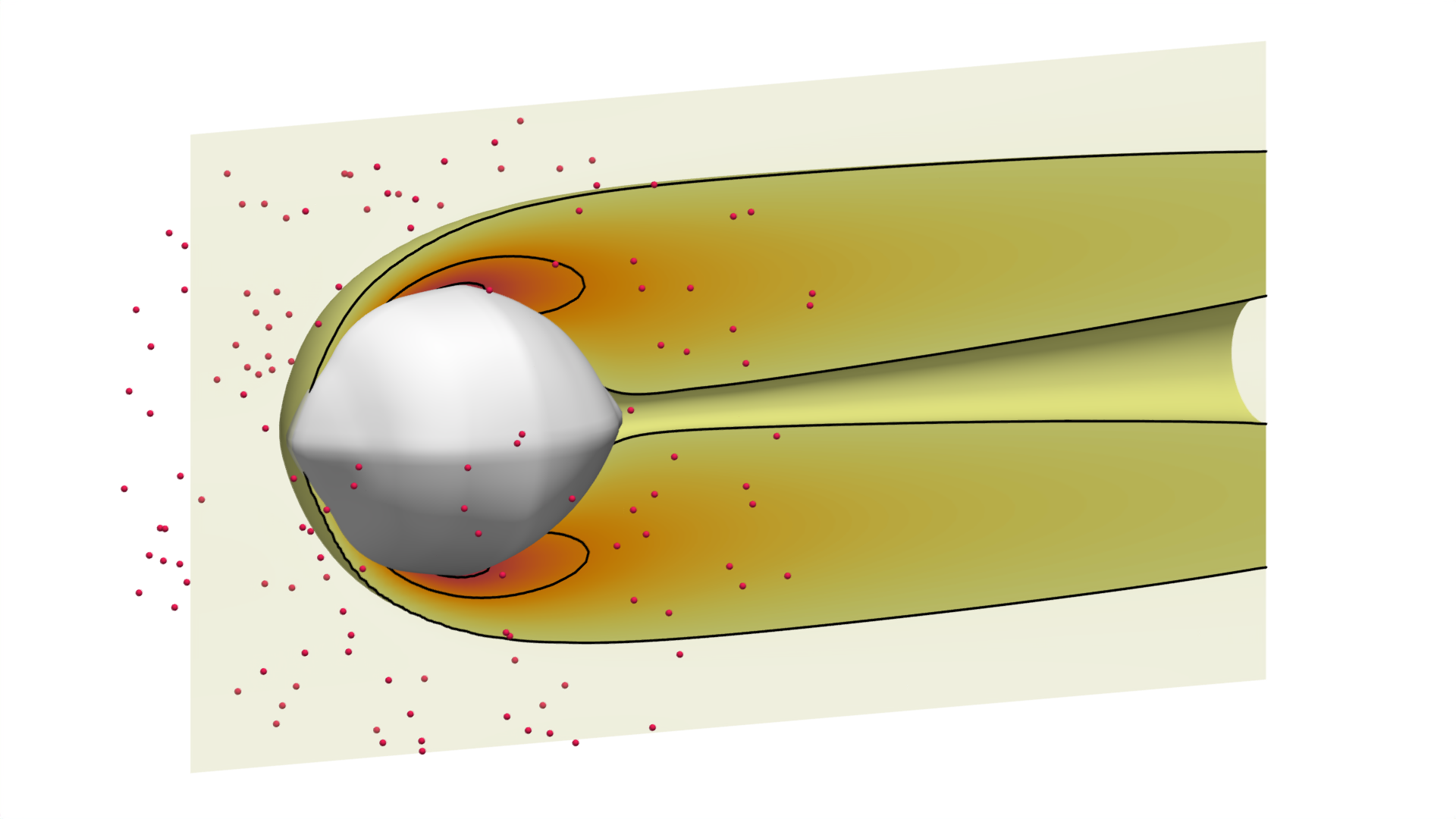}
      \put(7,47){\figbfpic{a}}
      \put(1,15){\rotatebox{90}{171 points}}
    \end{overpic}
    \begin{overpic}[width=7cm]{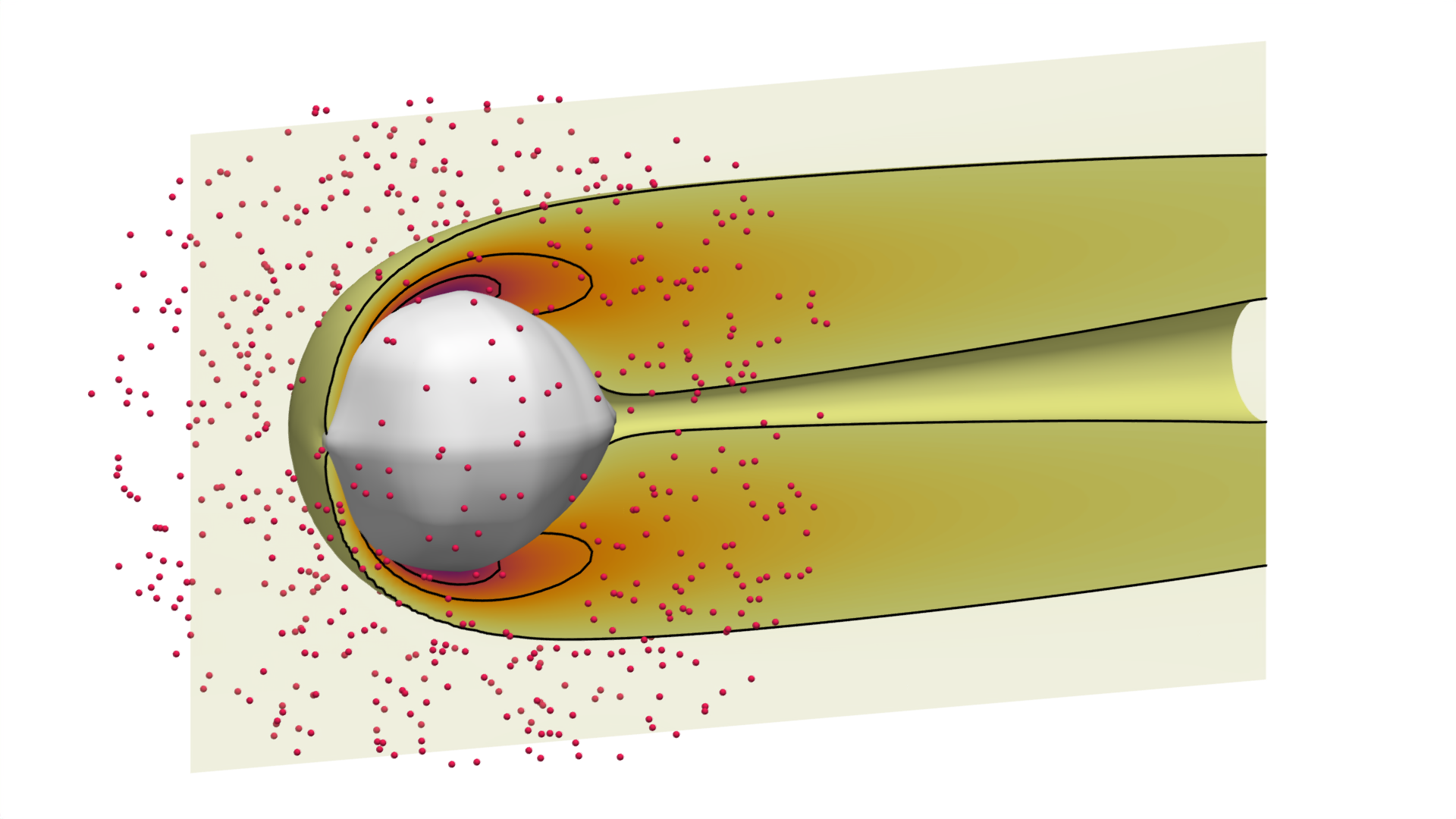}
      \put(7,47){\figbfpic{b}}
      \put(1,15){\rotatebox{90}{684 points}}
    \end{overpic}
    \begin{overpic}[width=7cm]{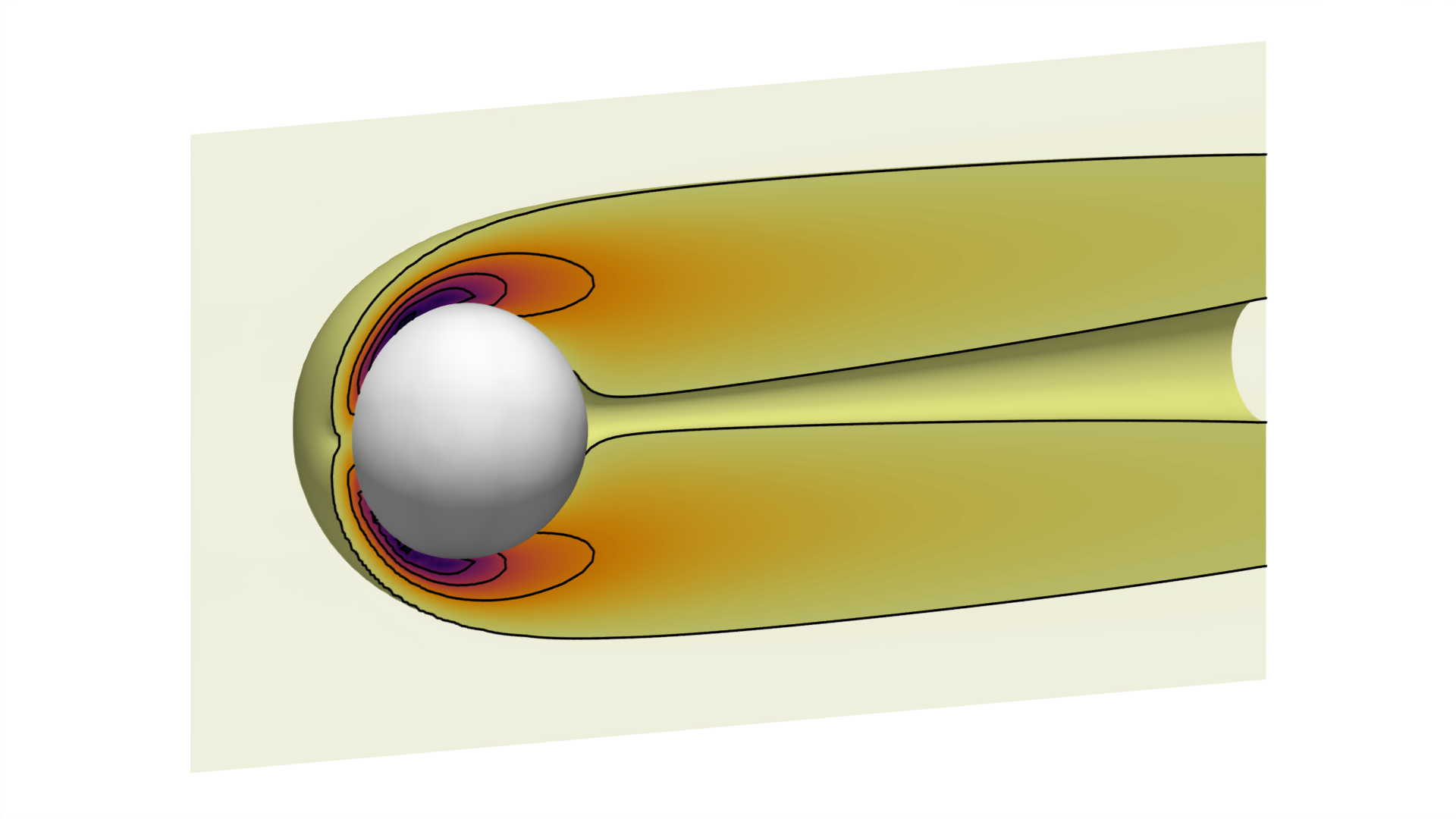}
      \put(7,47){\figbfpic{c}}
      \put(1,15){\rotatebox{90}{reference}}
    \end{overpic}
  \end{minipage}%
  \begin{minipage}[t]{0.15\textwidth}
    \includegraphics{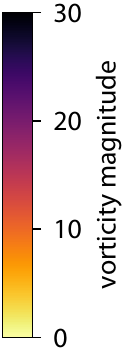}
  \end{minipage}%
  \begin{minipage}[t]{0.4\textwidth}
    \begin{overpic}{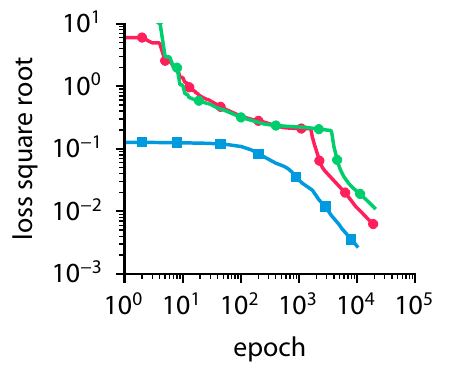}
      \put(2,78){\figbfpic{d}}
    \end{overpic}
    \begin{overpic}{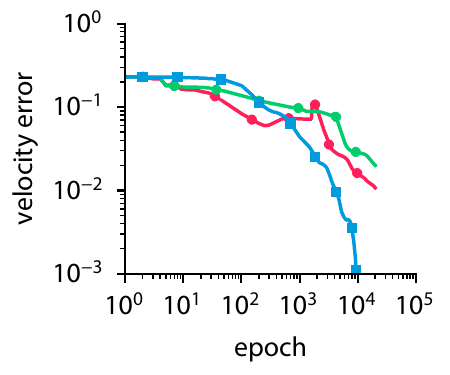}
      \put(2,78){\figbfpic{e}}
    \end{overpic}
    \begin{overpic}{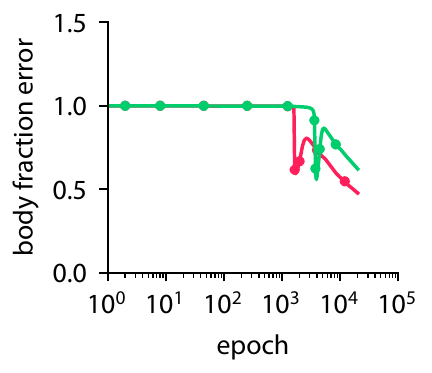}
      \put(2,82){\figbfpic{f}}
    \end{overpic}
  \end{minipage}

  \caption{%
    Inferring body shape from velocity measurements in three dimensions using mODIL.
    The reference shape is a sphere.
    (\figbf{a})~Body shape and contours of vorticity magnitude
    inferred from velocity in 171 points (red dots).
    (\figbf{b})~Body shape and contours of vorticity magnitude
    inferred from velocity in 684 points (red dots).
    (\figbf{c})~Reference body shape and contours vorticity magnitude.
    (\figbf{d},\figbf{e},\figbf{f})~Convergence history of mODIL with L-BFGS
    solving the inverse problem with 684 points~\key{111},
    inverse problem with 171 points~\key{122},
    and the forward problem~\key{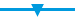},
    showing the square root of the loss function, RMS error in the velocity field,
    RMS error in the body volume fraction normalized by the
    reference volume.
  }
  \label{f_body3d_circle}
\end{figure*}

\begin{figure*}
  \centering
  \begin{minipage}[t]{0.4\textwidth}
    \begin{overpic}[width=7cm]{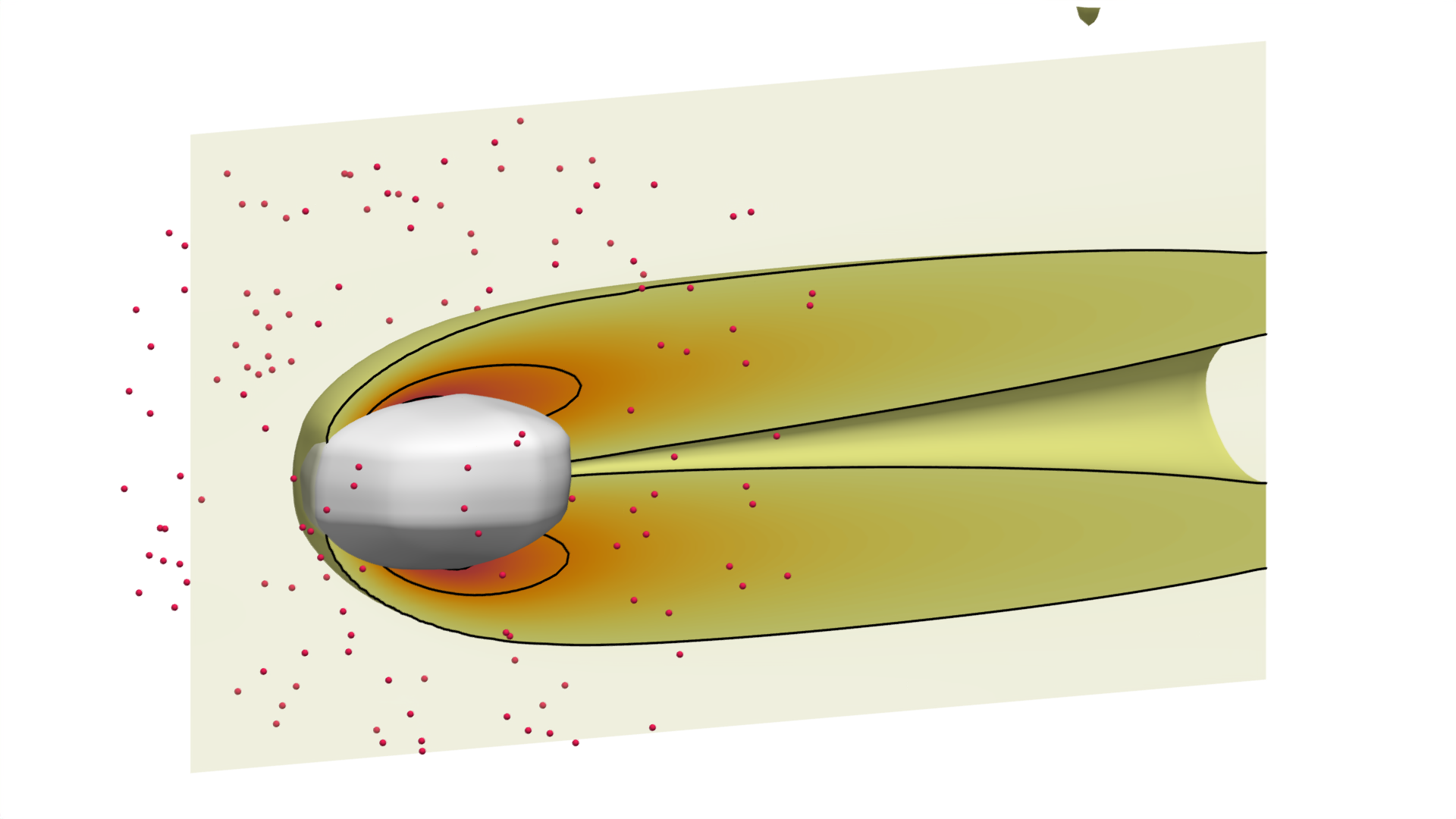}
      \put(7,47){\figbfpic{a}}
      \put(1,15){\rotatebox{90}{171 points}}
    \end{overpic}
    \begin{overpic}[width=7cm]{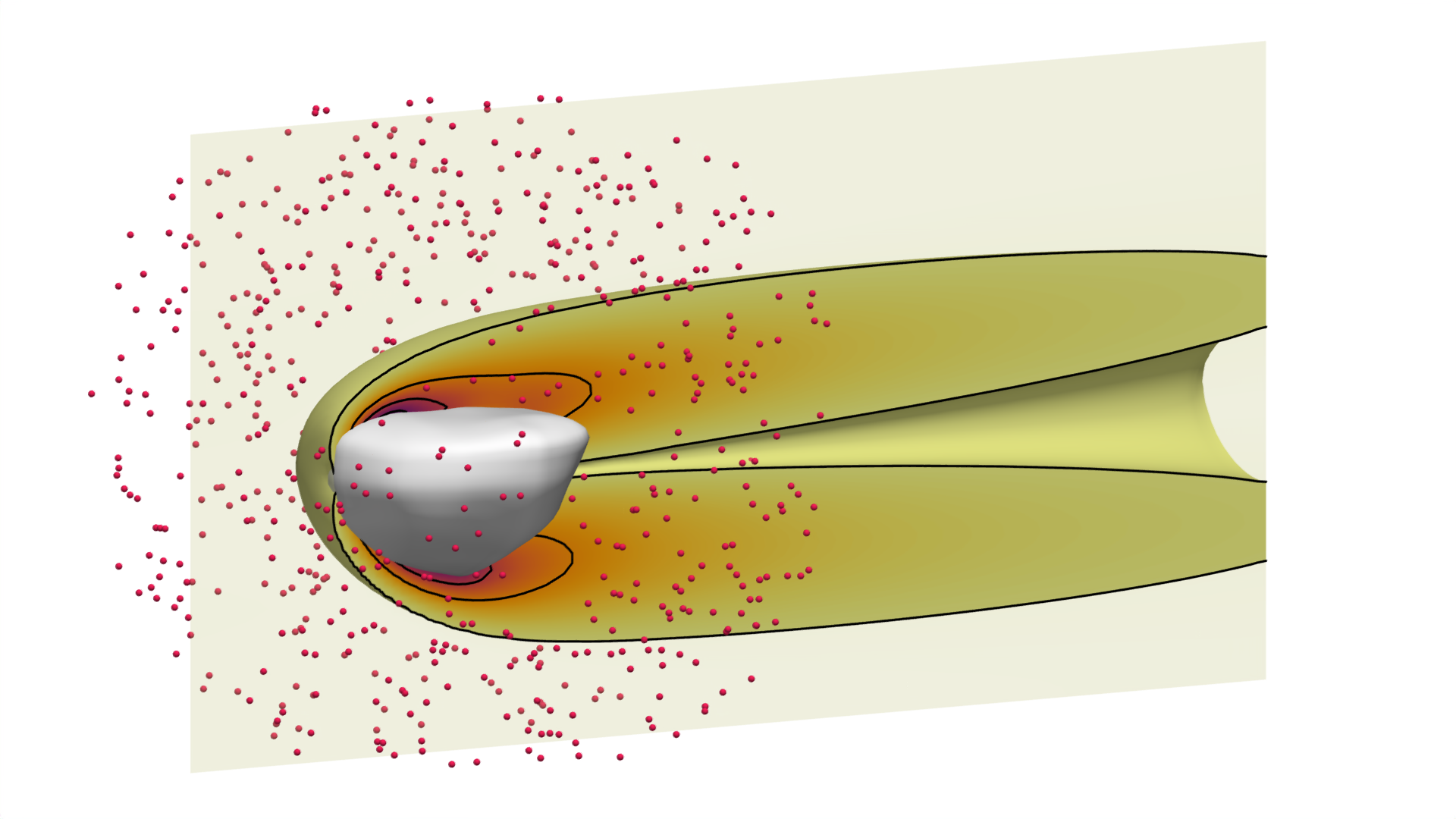}
      \put(7,47){\figbfpic{b}}
      \put(1,15){\rotatebox{90}{684 points}}
    \end{overpic}
    \begin{overpic}[width=7cm]{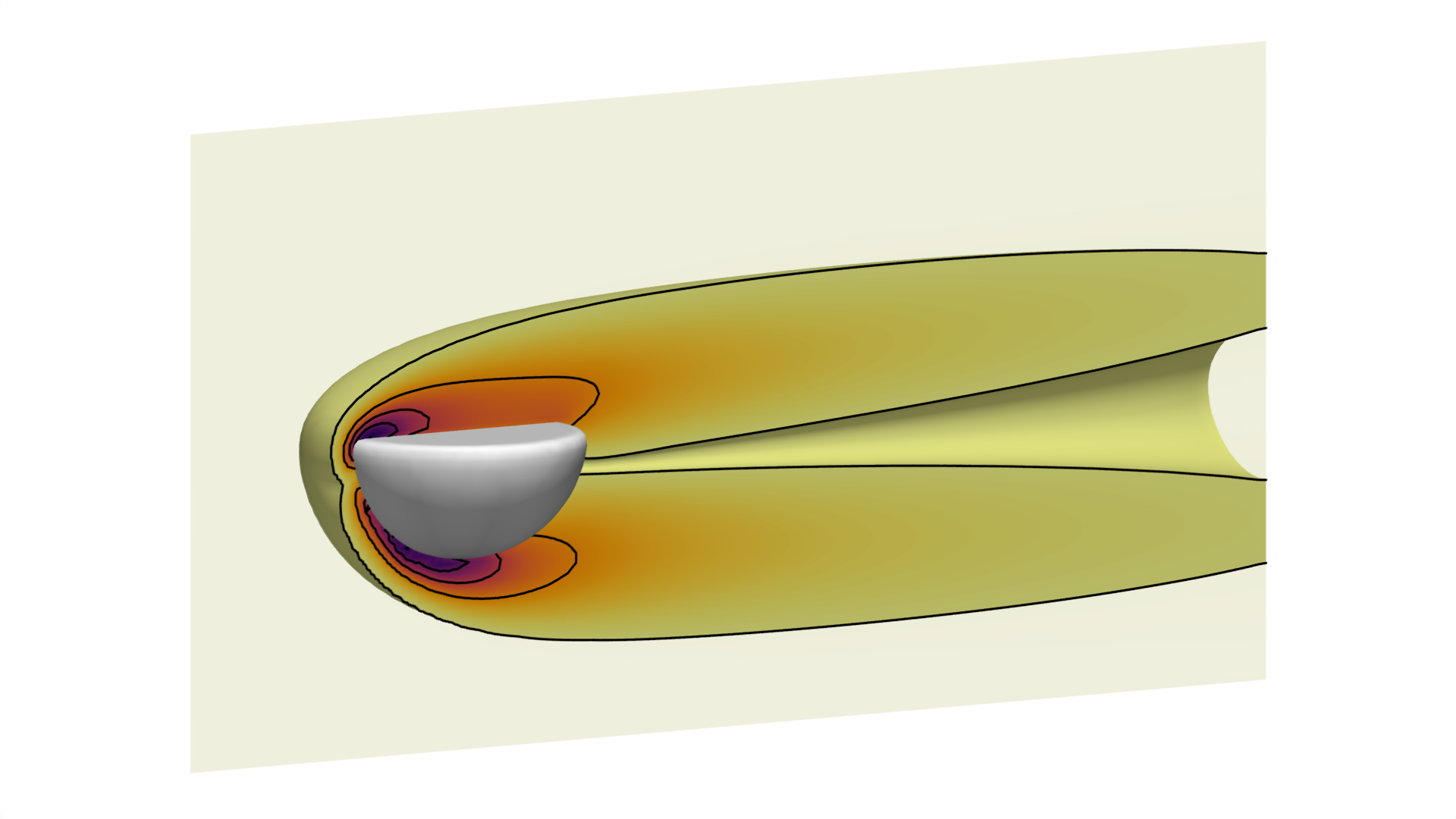}
      \put(7,47){\figbfpic{c}}
      \put(1,15){\rotatebox{90}{reference}}
    \end{overpic}
  \end{minipage}%
  \begin{minipage}[t]{0.15\textwidth}
    \includegraphics{body3d/cbar.pdf}
  \end{minipage}%
  \begin{minipage}[t]{0.4\textwidth}
    \begin{overpic}{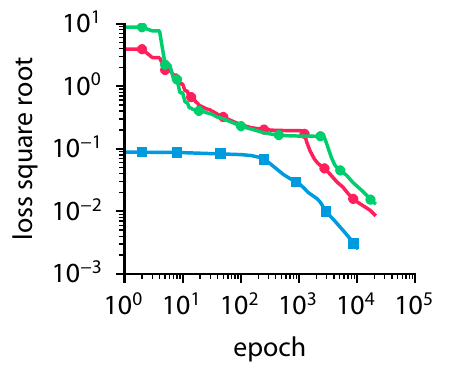}
      \put(2,78){\figbfpic{d}}
    \end{overpic}
    \begin{overpic}{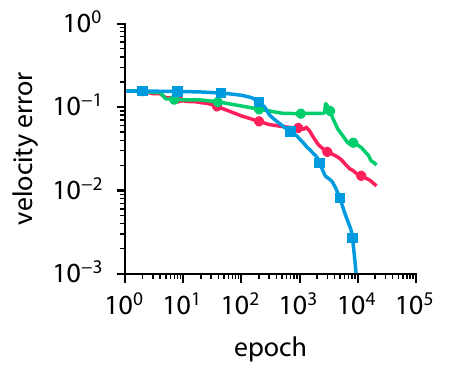}
      \put(2,78){\figbfpic{e}}
    \end{overpic}
    \begin{overpic}{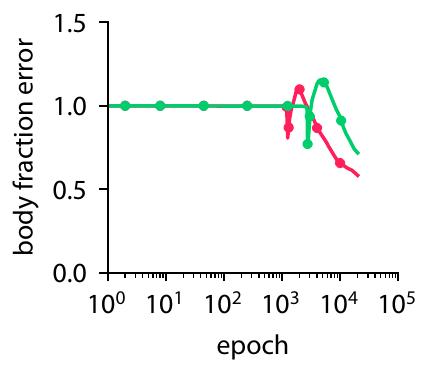}
      \put(2,82){\figbfpic{f}}
    \end{overpic}
  \end{minipage}

  \caption{%
    Inferring body shape from velocity measurements in three dimensions using mODIL.
    The reference shape is a hemisphere.
    (\figbf{a})~Body shape and contours of vorticity magnitude
    inferred from velocity in 171 points (red dots).
    (\figbf{b})~Body shape and contours of vorticity magnitude
    inferred from velocity in 684 points (red dots).
    (\figbf{c})~Reference body shape and contours vorticity magnitude.
    (\figbf{d},\figbf{e},\figbf{f})~Convergence history of mODIL with L-BFGS
    solving the inverse problem with 684 points~\key{111},
    inverse problem with 171 points~\key{122},
    and the forward problem~\key{133},
    showing the square root of the loss function, RMS in the velocity field,
    RMS error in the body volume fraction normalized by the
    reference volume.
  }
  \label{f_body3d_half}
\end{figure*}

\Cref{f_body3d_circle,f_body3d_half} show the results of the inference 
from 684 and 171 measurement points for two different bodies: a sphere and a hemisphere.
The sphere is centered at $(0.5,0.5,0.5)$ and has a radius of 0.2.
The hemisphere is an intersection of the sphere and the set of points $\{y<0.5\}$.
The convergence history includes the velocity error and the body fraction
error which are defined relative to the solution of the forward problem.
For both reference shapes, ODIL recovers a body shape
that qualitatively agrees with the reference,
although the relative error in the body fraction field amounts to 50\%,
so the inferred body volume is larger.
Using more measurement points for the inference reduces the error.
On a GPU Nvidia A100, the forward problem with a sphere
takes 53 minutes in total and $320~\text{ms}$ per epoch,
while the inverse problem takes 122 minutes in total
and $366~\text{ms}$ per epoch.
We note that solving the same inverse problem on a finer grid
of $257\times129\times129$ points takes $132$ minutes in total
and $400~\text{ms}$ per epoch.
Therefore, an eightfold increase in the number of grid points will lead
to a minor additional cost in the execution time of 8\%,
since the GPU operates more efficiently with larger arrays.

\section{Conclusion}
We have introduced the multigrid decomposition technique
that accelerates the convergence of gradient-based methods
for optimization problems that involve discrete fields on a grid.
The Multiresolution Optimization of the discrete loss (mODIL) is 
based on a hierarchy of successively coarser grids and extends
significantly the capabilities of the recently introduced
ODIL (Optimizing a DIscrete Loss) framework~\cite{karnakov2022optimizing}.
The multigrid decomposition represents a discrete field
as a sum of fields interpolated from all grid levels,
increasing the number of parameters.
This technique addresses the issue of locality of gradient-based optimizers
by extending the domain of dependence of each scalar parameter
so that information can propagate through the grid faster.
Gradients of the resulting loss function can be computed using automatic
differentiation, making its implementation straightforward.

ODIL introduces a new modality in solving fluid mechanics problems.
Notable large scale  simulations of the Navier Stokes equations
(examples of recent works include \cite{falcucci2021extreme,karnakov2022computing,andersson2023inferno})
are based on  of forward time marching solutions using supercomputers
with hundreds of GPUs and billions or trillions~\cite{rossinelli201311} of computational elements.
Our current implementation is limited to computations on one GPU on a grid of $\mathcal{O}(4\times 10^6)$ points. 
Extending our implementation to computations using multiple GPUs is the subject of ongoing work.
We note that ODIL  provides a  "one shot" solution instead of time stepping for solving fluid mechanics problems.
In its present form ODIL requires large memory resources and "replaces" time stepping with iterations of an optimization.

We demonstrate the effectiveness of mODIL on a variety of forward and inverse
problems, including flow reconstruction from sparse measurements.
The multigrid formulation takes up to 10x fewer iterations to achieve the same
error, better avoids local minima, and results in more regular convergence
behavior. Our results suggest that mODIL 
represents a state of the art method for solving 2D and 3D inverse problems in fluid mechanics.
Work is underway to extend mODIL to inverse problems across different scientific fields.

\subsection*{Code availability}

The software implementation of the method is available
at~\url{https://github.com/cselab/odil} along with examples and
instructions to reproduce the results.

\subsection*{Data Availability}
The datasets generated during and/or analysed during the current study are available from the corresponding author on reasonable request.

\subsection*{Author contributions}
Petr Karnakov developed the multigrid decomposition technique and programmed the software.
Sergey Litvinov and Petros Koumoutsakos assisted in the formulation of the research and on the writing of the manuscript.

\bibliography{mainbib}

\end{document}